\documentclass[sn-mathphys,Numbered,iicol]{sn-jnl}

\usepackage{graphicx}%
\usepackage{multirow}%
\usepackage{amsmath,amssymb,amsfonts}%
\usepackage{amsthm}%
\usepackage{mathrsfs}%
\usepackage[title]{appendix}%
\usepackage{xcolor}%
\definecolor{eminence}{RGB}{255,0,0}
\usepackage{textcomp}%
\usepackage{manyfoot}%
\usepackage{booktabs}%
\usepackage{algorithm}%
\usepackage{algorithmicx}%
\usepackage{algpseudocode}%
\usepackage{listings}%
\usepackage{soul} 
\usepackage{tabularx}
\usepackage{float}
 
 \theoremstyle{thmstyleone}%
\theoremstyle{thmstyletwo}%

\theoremstyle{thmstylethree}%

\begin{document}

\title[Article Title]{Modeling Dislocation Dynamics Data Using Semantic Web Technologies}




	
\author[1]{\fnm{Ahmad Zainul} \sur{Ihsan}}\email{a.ihsan@fz-juelich.de}
\author[1]{\fnm{Said} \sur{Fathalla}}\email{s.fathalla@fz-juelich.de}

\author*[1,2]{\fnm{Stefan} \sur{Sandfeld}}\email{s.sandfeld@fz-juelich.de}

\affil*[1]{\orgdiv{Institute of Advanced Simulation--Materials Data Science and Informatics (IAS-9)}, \orgname{Forschungszentrum Jülich}, \country{Germany}}
\affil[2]{\orgdiv{Faculty of Georesources and Materials Engineering}, \orgname{RWTH Aachen University}, \country{Germany}}


\abstract{
Research in the field of Materials Science and Engineering focuses on the design, synthesis, properties, and performance of materials. 
An important class of materials that is widely investigated are crystalline materials, including metals and semiconductors. 
Crystalline material typically contains a distinct type of defect called ``dislocation''. 
This defect significantly affects various material properties, including strength, fracture toughness, and ductility. 
Researchers have devoted a significant effort in recent years to understanding dislocation behavior through experimental characterization techniques and simulations, e.g., dislocation dynamics simulations.
This paper presents how data from dislocation dynamics simulations can be modeled using semantic web technologies through annotating data with ontologies.
We extend the already existing Dislocation Ontology by adding missing concepts and aligning it with two other domain-related ontologies (i.e., the Elementary Multi-perspective Material Ontology and the Materials Design Ontology) allowing for representing the dislocation simulation data efficiently. 
Moreover, we show a real-world use case by representing the discrete dislocation dynamics data as a knowledge graph (DisLocKG) that illustrates the relationship between them. 
We also developed a SPARQL endpoint that brings extensive flexibility to query DisLocKG. 
}

\maketitle

\section{Introduction}
\label{sec1/introduction}

Plastic deformation in metals and other crystalline materials can be attributed to a one-dimensional lattice defect type known as dislocation. The concept of the dislocation was introduced in the 1930s by 
Taylor~\cite{taylor1934a} and Polanyi~\cite{polanyi1934}. 
Dislocations determine mechanical properties of materials, such as strength, hardness, and ductility. 
For instance, materials engineers have discovered the strengthening mechanism of crystalline materials by studying the relationship between dislocation motion and the mechanical behavior of metals~\cite{callister2018}.  
By controlling the motion of dislocations in crystalline materials, materials engineers can build, for example, an airplane turbine blade that can withstand an operation temperature of $\sim1000^{\circ}C$ and creep deformation due to centrifugal forces while the turbine is rotating~\cite{murakumo2004creep}. 
Significant efforts have been made to understand dislocation systems using dedicated microscopy techniques and simulation methods. 
These simulation methods along with other techniques have been created to predict dislocation evolution.

In recent years, data-driven approaches have brought new methods and tools for analyzing and understanding the evolution of dislocation systems ~\cite{govind2023deep,bertin2023accelerating,zhang2022data,yang2020learning,song2021data,salmenjoki2018machine}. 
This intensely transforms the Materials Science and Engineering (MSE), combining simulations, data mining, and experiments, making the digital transformation possible~\cite{Prakash2018_PrM55,EMMC-MatDig}. 
However, a digital transformation without being supported by the appropriate data infrastructure often ends up with isolated and inaccessible data repositories, the so-called ``{data silos}''. 
In this regard, materials informatics plays a significant role in materials science research to overcome the data silos problem. 
This is because materials informatics combines two discourses of materials science and information technologies to tackle the major problems in materials science, such as data management and analysis. 
Moreover, it helps to develop intelligent systems to, e.g., explore materials, find novel materials properties, or study the behavior of a specific materials phenomenon.

To fully understand the behavior of materials and in particular dislocations, aspects from different length scales need to be considered. 
This variety of length scales makes the knowledge representation of systems of dislocations challenging, even though this has yet to be perceived as a significant research hindrance in materials science. 
Generally, the schematic representation of knowledge, i.e. the representation through ontologies, can significantly boost data management and analysis; it helps to extract knowledge from data.
Ontologies also allow the domain knowledge to be machine-understandable, meaning that machines can read and interpret this knowledge efficiently. 
Furthermore, it has become an essential part of achieving FAIR (Discoverable, Accessible, Interoperable, and Reusable) data~\cite{wilkinson2016}.

This paper presents how Discrete Dislocation Dynamics (DDD) data can be enriched using Semantic Web technologies, such as the Resource Description Framework (RDF)~\cite{mcbride2004resource}, the Web Ontology Language (OWL)~\cite{bechhofer2004owl} and SPARQL~\cite{perez2006semantics}. 
The first step we have taken is to adapt and extend the dislocation ontology (DISO)~\cite{IhsanDISO2023, ihsan2021_1} so that it can model various concepts and relationships in the DDD domain.
The adaption includes adding missing concepts, improving class definition, exploring additional relationships between concepts, and finally aligning it with other domain-related ontologies, including the Elementary Multi-perspective Material Ontology (EMMO) and the Materials Design Ontology (MDO).
This allows for representing the dislocation simulation data efficiently. 
DISO is one of the Dislocation Ontology Suite (DISOS)\footnote{\url{https://purls.helmholtz-metadaten.de/disos}} ontologies that represent the concepts and relationships of linear defects in crystalline materials.
In fact, DISOS comprises several modules describing materials scientific concepts, representations of dislocations, and different simulation models in the dislocation domain. 
The adapted version of DISO is developed and maintained in the DISOS GitHub repository.
The ontology is available in several RDF serializations via a persistent identifier (i.e., {\url{https://purls.helmholtz-metadaten.de/disos/diso}}) provided by PIDA (Persistent Identifiers for Digital Assets)\footnote{\url{https://purls.helmholtz-metadaten.de/}}.
PIDA employs content negotiation~\cite{berrueta2008cooking} to serve different versions of the ontology (i.e., the HTML documentation or an RDF representation) via its IRI. 
DISO has been syntactically validated by the W3C RDF validation service\footnote{\url{https://www.w3.org/RDF/Validator/}} to conform with the W3C RDF standards. 
The documentation of the ontology is available via its IRI.

The next step after adapting  the ontology is to annotate the data gathered from multiple DDD simulations with the adapted version of DISO resulting in a knowledge graph (DisLocKG) of DDD data (more details can be found in \autoref{sec6/real-world-use-case}).
This knowledge graph connects DDD data concepts via the relationship between them, thus enabling machine actionability \cite{loibl2020procedure}, allowing for semantic querying, inferring implicit knowledge that does not exist, and ensuring data consistency and integrity.
The objective is to convert the unstructured DDD data to linked data with dereferenceable IRIs that adheres to W3C standards and best practices. 
This will enable not only reasoning about the dislocation data but also to integrate it into other MSE-related fields. 
We have made DisLocKG publicly available via its GitHub repository\footnote{\url{https://purls.helmholtz-metadaten.de/dislockg}}.


\section{Related work}
\label{sec2/related_work}
Over the past few years, many researchers gave a particular attention on developing ontologies to represent scientific data in different fields of science, such as physics \cite{say2020semantic}, agriculture \cite{hu2011agont}, and pharmaceutical science \cite{say2020ontology}. 
Specifically in the MSE field, several efforts have been up in creating ontologies representing materials-related notions or semantically presenting actual materials data as knowledge graphs. This section will discuss related works of knowledge graphs with and without semantic web technology (including RDF, OWL, and SPARQL).
Two examples from the latter are the \textit{Propnet} Knowledge Graph~\cite{mrdjenovich2020propnet} and the Materials Knowledge Graph (\textit{MatKG})~\cite{zhao2021knowledge}.

The Propnet is a knowledge graph enhancing materials properties data from the Materials Project~\cite{jain2013} Repository\footnote{\url{https://next-gen.materialsproject.org}}. 
It augments base properties data (e.g., lattice, basis, chemical formula, band gap, and total energy) resulting from the ab-initio calculation into derived properties, e.g., Debye temperature, bulk modulus, and shear modulus. 
The workflow and input for generating the augmented data are subsequently stored in the knowledge graph.

On the other hand, the MatKG stores metadata from over 2.9 million materials science articles. 
This metadata includes abstracts, titles, keywords, and author data (e.g., name, email, affiliation, and ORCID). 
By accessing the MatKG, we can retrieve information such as the milestones of a material developed by multiple authors.

\citet{ashino2010} have developed a ``Materials Ontology'' which is an ontology describing substances, processes, environments, and properties. 
This ontology also has been used to exchange data between three different thermal property databases. 
%

In the solid-state physics domain, \citet{li2020} have developed the \textit{Materials Design Ontology} (MDO) which is an ontology covering knowledge in the field of materials design, e.g., with regards to ab-initio methods.
MDO is used to represent materials' data related to ab-initio calculations over disparate materials data repositories as RDF triples.
At the time of writing the paper, a total of $\approx 4.3$K RDF triples have been collected in their repository\footnote{\url{https://github.com/LiUSemWeb/materials-design-ontology}}. 
While the work is related to the representation of crystalline material by means of the crystal structure, MDO does not represent data related to crystalline defects.

Another effort in the experimental materials science community that  uses semantic web technologies is the NanoMine Knowledge Graph\footnote{\url{http://nanomine.org/}}~\cite{mccusker2020nanomine}. 
It is a knowledge graph for polymer nanocomposite materials integrating diverse data from more than 1,700 polymer nanocomposite experiments. 
Moreover, the authors of the NanoMine knowledge graph have developed the NanoMine ontology\footnote{\url{https://github.com/tetherless-world/nanomine-ontology}}, which is a backbone ontology to describe polymer nanocomposite experiments.



In conclusion, it is evident that even though several efforts and groups utilizing the semantic web in various MSE-related fields have progressed significantly, work for semantically representing dislocations simulation data is still missing. 
We believe that this work is the first attempt at creating a knowledge graph in an MSE-related domain that deals with data governing details of linear crystallographic defects, i.e. dislocation data. 
As a result, the unstructured dislocation data is transformed into linked data with dereferenceable IRIs using persistent URLs, adhering to W3C standards and best practices. 
This enables not only to annotation of dislocation data by an ontology but also to integration of dislocation data into other MSE-related fields. 

\section{Description of the Domain}
\label{sec3/domain-description}
This section briefly describes the relevant notions and concepts of line defects within the crystalline materials domain. 

\subsection{Representation of Crystalline Materials}
\label{sec3/subsec/crystalline-materials}
\begin{figure}[htbp]
	\centering
	\includegraphics[width=.8\linewidth]{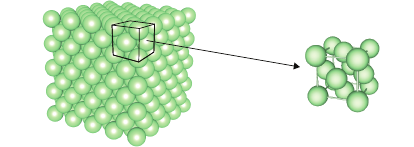}
	\caption[]{The crystal structure of face-centered cubic comprises an aggregate of atoms within one unit cell in crystalline materials.}
	\label{fig/sec3:crystal-structure}
\end{figure}

Most of the metals and metallic materials have a crystalline structure, which implies that the
atoms are arranged in a periodic structure with a high degree of symmetry.
This periodic arrangement is at the basis of the \emph{crystal structure} model, idealizing the physical concept of crystalline materials. For example, in \autoref{fig/sec3:crystal-structure} atoms are shown in an idealized manner as small spheres.
The crystal structure is represented by the \emph{lattice} together with a \emph{motif}: the lattice is a mathematical concept of an infinite, repeating arrangement of points in space (3D), in a plane (2D), or on a line (1D), in which all points have the same surrounding and coincide with atom  positions. 
The motif (or base) consists of an arrangement of chemical species, which can be atoms, ions, or molecules in crystalline materials. 
By putting a motif of one or more atoms at every lattice point, the crystal structure can be represented.

\begin{figure}[tbp]
    \centering
	\includegraphics[width=0.7\linewidth]{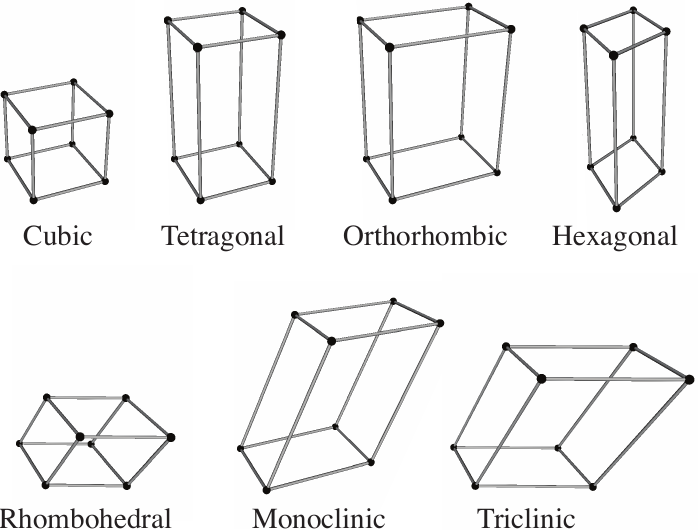}
	\caption[]{The seven crystal systems. These seven crystal systems are also seven primitive Bravais lattices. Each of them only corresponds to a single lattice point.}
	\label{fig/sec3:crystal-system}
\end{figure}

It is now possible to identify the smallest atom pattern that can be repeated along all spatial directions to cover the entire structure.
This pattern is called a \emph{unit cell}, shown as the black cube in \autoref{fig/sec3:crystal-structure}. 
The lattice parameters of the unit cell consist of the angles between the edges and the edge lengths. 
\begin{figure}[tbp]
	\includegraphics[width=.9\linewidth]{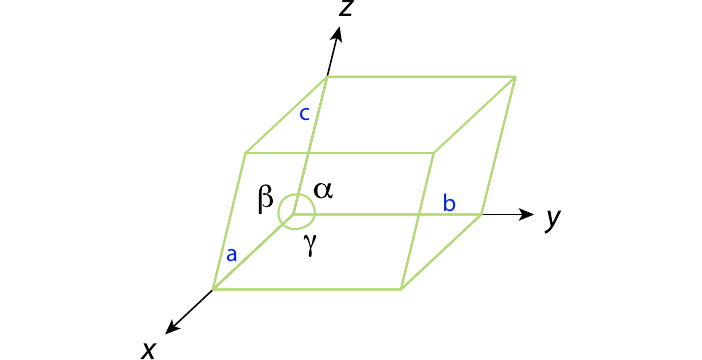}
	\caption[]{The geometry of a unit cell is exactly defined through the three lengths, $a$, $b$, and $c$,  and the three angles $\alpha$, $\beta$, and $\gamma$.}
	\label{fig/sec3:unit-cell}
\end{figure}
\autoref{fig/sec3:unit-cell} shows the six lattice parameters needed to characterize the unit cell: three lengths ($a, b, c$) and three angles ($\alpha, \beta, \gamma$). 
These parameters also constitute the basis vectors in the crystal coordinate system; they are not necessarily mutually perpendicular. 
Unit cells are often classified into a systematic based on the lattice parameters (cf. \autoref{fig/sec3:crystal-system}). 
For instance, the cubic system has $a=b=c,\; \alpha=\beta=\gamma=90^\circ$ and the orthorhombic system has $a\neq b\neq c,\; \alpha = \beta = \gamma = 90^\circ$.
Seven crystal systems are often ordered according to the increasing symmetry: cubic, tetragonal, orthorhombic, hexagonal, rhombohedral, monoclinic, and triclinic.

\begin{figure}[htbp]
	\includegraphics[width=\linewidth]{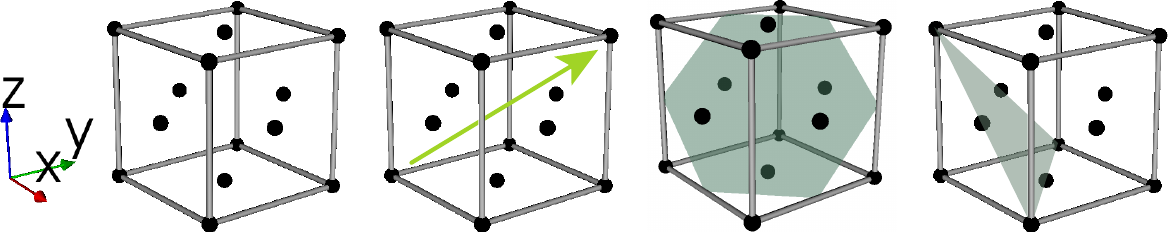}
	\caption[]{On the very left, a unit cell is shown that corresponds to a "face-centered cubic" structure. The black points indicate the lattice points, i.e. the positions of the atoms. Second from left: a direction vector that connects to these lattice points. Lastly, two different lattice planes are shown.}
	\label{fig/sec3:lattice-plane}
\end{figure}

In the unit cell, we can also define \emph{lattice points}, \emph{lattice directions}, and \emph{lattice planes}:
A lattice consists of lattice points where the atoms, ions, or molecules are located (the leftmost cube in  \autoref{fig/sec3:lattice-plane}). 
The vector position of lattice points, $\overrightarrow{R}$, is described by the equation 
\begin{equation}
  \mathbf{\overrightarrow{R}} = n_1\mathbf{a} + n_2\mathbf{b} + n_3\mathbf{c}\;,
  \label{eq/lattice-point}
\end{equation}
where $n_i$ are arbitrary integers and $\mathbf{a}, \mathbf{b}, \mathbf{c}$ are basis vectors (pointing along the axes in \autoref{fig/sec3:unit-cell}) derived from the lattice parameters. 
As illustrated in \autoref{fig/sec3:lattice-plane}, a lattice direction or lattice vector is a vector connecting two lattice points, whereas a lattice plane forms an infinitely stretched plane (characterized through a plane normal) that cuts through lattice points such that a regular arrangement of lattice points in the plane occurs.


\subsection{Description of Linear Defects}
\label{sec3/subsec/linear-defects}

\begin{figure}[tbp]
    \centering
   \includegraphics[width=0.8\linewidth]{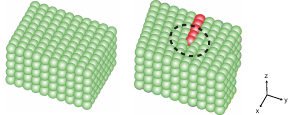}
   \caption[]{The left figure shows the perfect order of atoms in a crystalline material. The right crystal contains a dislocation that destroys the local order of the crystal structure. Note that red-colored atoms only show the `irregularity'', and are not different atom types.}
   \label{fig/sec3:perfect-defect}
\end{figure}

In crystalline materials, atoms are not always arranged or positioned perfectly. 
Typically, different kinds of crystallographic defects lead to disruption of the local order in a material (in addition to thermal fluctuations affecting the atomic positions). 
A common type of such defect is the dislocation, which causes a strongly localized, tube-like region of disorder (illustrated by the dashed circle in the right panel of \autoref{fig/sec3:perfect-defect}; this tube-like region stretches along the $z$-direction). 
This region contains the highly disordered dislocation core at the center. Further away from the dislocation core, the perfect lattice structure is restored, even though there is now a row of atoms shifted into the new position as indicated by the red spheres in \autoref{fig/sec3:perfect-defect}.

\begin{figure}[tbp]
    \centering
  \includegraphics[width=.8\linewidth]{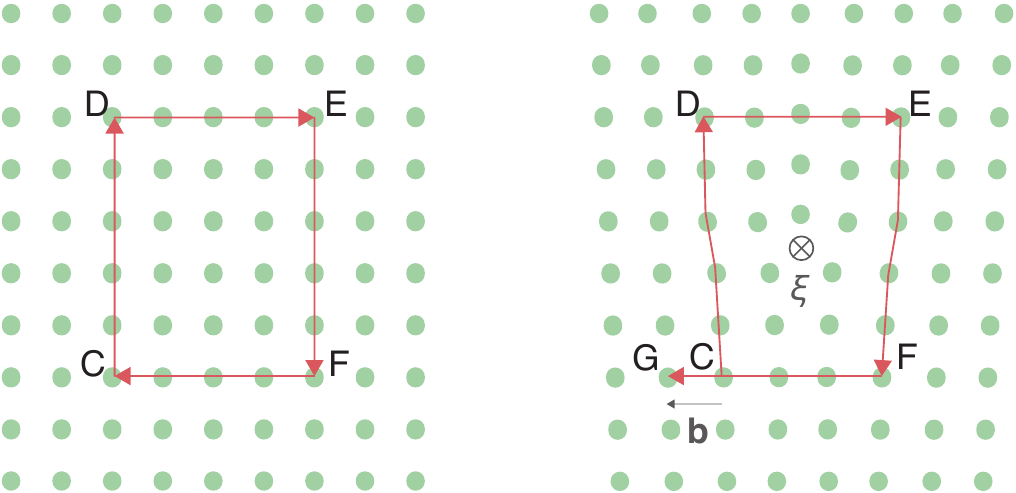}
  \caption[]{The Burgers circuit in the crystalline materials. While the left panel has the Burgers circuit in the perfect crystal material. The right panel has the Burgers circuit around the dislocation in the crystalline materials. Due to the closure failure in the defective crystalline materials, we can define the Burgers vector, $\mathbf{b}$.}
  \label{fig/sec3:burgersvector}
\end{figure}

The ``Burgers vector'' of a dislocation can be defined through the ``Burgers circuit'', as shown in \ref{fig/sec3:burgersvector}.
A Burgers circuit is an atom-to-atom path that is closed in a perfect crystal (left panel of the figure). The length of the path is given as multiple of the atomic distances in two directions. In the presence of a dislocation, a path of the same lengths would not be closed (right panel of the figure). 
The step-by-step procedure is as follows:
We define a reference point \textit{C} as the start point of the path. The 
line sense of the path is given by $\boldsymbol{\xi}$ assuming the "right-hand convention" (the thumb points along the vector $\boldsymbol{\xi}$ into the picture plane and the other fingers curl around that vector; their fingertips indicate the direction of the path).
The symbol $\otimes$ in the figure indicates that the vector $\boldsymbol{\xi}$ points into the picture plane.
In the perfect crystal, this circuit goes up five atoms from the reference point, four atoms to the left, five atoms down, and four atoms to the left to close a circuit at point \textit{C} again. 
With the same reference point and the same number of atomic spacings as in the circuit, the perfect crystalline material can be used with the crystal  containing a dislocation. 
However, the circuit does not end at the same point as in the perfect crystal, but rather at point \textit{G} as shown in the right panel of \autoref{fig/sec3:burgersvector}.
The vector connecting the start point \textit{C} with the finish point \textit{G} is the Burgers vector $\mathbf{b}$. 

There are two fundamentally different types of dislocations: ``screw'' and ``edge'' dislocation. 
A screw dislocation has a line sense parallel to its Burgers vector, $\boldsymbol{\xi}||\boldsymbol{b}$, whereas for an edge dislocations the line sense is perpendicular to its Burgers vector, $\boldsymbol{\xi}\perp\boldsymbol{b}$. Thus, \autoref{fig/sec3:burgersvector} shows an edge dislocation.

In reality, screw and edge dislocations are extreme cases, while the most general case of a dislocation type in a crystalline material is a ``mixed dislocation.''
This is a dislocation with the line sense, $\boldsymbol{\xi}$, neither parallel nor perpendicular to its Burgers vector, $\mathbf{b}$.

Since the atoms around dislocations are not positioned at the perfect lattice points, the lattice is distorted near a dislocation. 
This distortion results in a stress field in the crystalline material around the dislocation which is the reason why dislocations move: they try to minimize energy. 
In the context of plastic deformation, a dislocation is defined as the boundary of a slipped area within which atoms are displaced by the size of an elementary unit translation given by the Burgers vector. 

\begin{figure}[htbp]
   \includegraphics[width=\linewidth]{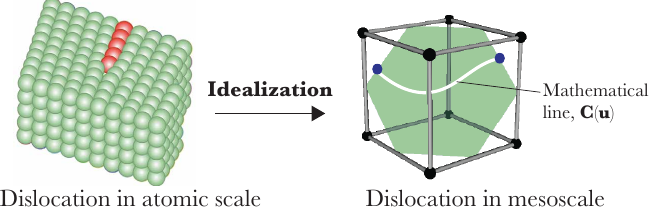}
   \caption[]{The idealization represents the dislocation in the mesoscale. Here, the individual atoms are no longer visible. This idealization reduced the tube-like defect ``region'' to a mathematical line. Note that the line on the right does not correspond to the dislocation in the left (this would be a vertical, straight line).}
   \label{fig/sec3:idealization-dislocation}
\end{figure}

In materials science, often the question arises on which ``granularity level'' a dislocation should be defined. Clearly, if we are interested in phenomena on the nanometer scale then we should resolve individual atoms (e.g., through high-resolution transmission electron microscopy or molecular dynamics simulations). 
When taking the \emph{mesoscopic} perspective, typically the individual atoms can not be seen anymore and are not of interest (as, e.g., done through regular transmission electron microscopy or dislocations dynamics simulations). However, the dislocation line itself is still observable: the tube-like defect ``region'' is reduced to an idealized mathematical line, as demonstrated in the right panel of \autoref{fig/sec3:idealization-dislocation}.
Therefore, the transition from the atomic scale to the mesoscale requires a conceptual and mathematical idealization that significantly reduces the amount of information.
These idealizations require to be accompanied by further details and definitions from the atomic scale, including the crystal structure, the lattice, the lattice plane, and the lattice direction information, all of which have an impact on the dislocation's motion.
For example, the motion of a dislocation line through a crystal is constrained to a specific crystallographic plane or lattice plane. 
Thus, it still requires crystallographic information, even though the ``defect region'' is now only represented as a mathematical line. 
These two different levels of information require particular attention when designing the dislocation ontology.

\begin{figure}[htbp]
   \centering
   \includegraphics[width=0.6\linewidth]{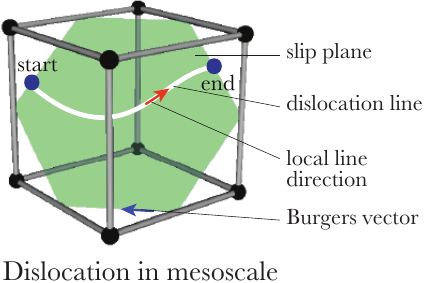}
   \caption[]{Depiction of the mathematical dislocation line on the mesoscale as a mathematical object that has start and end points. The object is characterized by the Burgers vector and the line sense. Furthermore, the dislocation motion is constrained by the slip plane.}
   \label{fig/sec3:lattice-plane-dislocation}
\end{figure}

The particular crystallographic or lattice plane constraining the dislocation motion is called the \emph{slip plane} (see the green plane in \autoref{fig/sec3:lattice-plane-dislocation}). 
There are specific \emph{slip directions}, which are lattice directions along which plastic deformation occurs within the slip plane, given by the Burgers vector. 
A \emph{slip system} is a set of slip planes with the same unit normal vector and the same slip direction. 
Thus, the unit normal vector and the slip direction or the Burgers vector (where the latter is not a unit vector) determine the slip system.

The mathematical representation of a mesoscale dislocation, as shown in \autoref{fig/sec3:lattice-plane-dislocation}, is an oriented curve with a start point and an endpoint. 
The local line orientation changes along the line, while the Burgers vector is constant for each point. 
Since the dislocation is a directed curve, it has a line sense. 
Unlike the local line orientation, it is a property of the whole line.


Various computational and experimental techniques are leveraged to predict and observe dislocations in crystalline materials, some of which were already mentioned above. 
For instance, high-resolution transmission electron microscopy (or field ion microscopy) is used on the atomic scale to image the arrangement of atoms. 
On the mesoscale, the focus is on examining the characteristics of individual dislocations and analyzing the distribution, arrangement, and density of dislocation in materials.
Transmission Electron Microscopy (TEM) and Discrete Dislocation Dynamics (DDD) simulations  are techniques for investigating these properties and simulating the dislocation behaviour, respectively.

TEM is a microscopy technique that generates a highly-magnified image of a material specimen. 
This technique involves an electron beam passing through the specimen and several lenses.
In strongly simplified terms, if the electron beam hits an atom, then it is deflected. 
As a result of the deflection, the intensity of the transmitted beam is reduced, and the intensity of the diffracted beam is increased. 
The dislocation can be seen as a dark line in such a bright-field image. 

Above, it was already mentioned that the displaced atoms around a dislocation result in 
stresses, because the atoms are no longer in their preferred equilibrium position. Dislocations move in such stress fields
which are mainly described by the governing equations of (linear) elasticity theory. 
DDD simulations employ mathematical lines (polygons or splines) to represent dislocations, which are moved based on elastic interactions and further ``local rules''. 

\begin{figure}[htbp]
    \centering
   \includegraphics[width=0.95\linewidth]{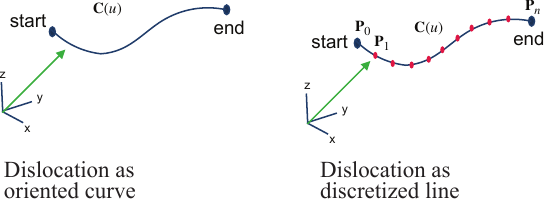}
   \caption[]{The discretization of dislocation to a numerical representation. The oriented curve dislocation line shown in the left panel is discretized into a number of segments shown in the right panel.}
   \label{fig/sec3:discretized-dislocation}
\end{figure}

The numerical schemes used in DDD simulations require to numerically discretize the mathematical line, e.g., by a number of straight line segments.
The discretization steps are illustrated in  \autoref{fig/sec3:discretized-dislocation}. 
Further details can be found in~\citep{ghoniem1999}. 
The discretization process can be easiest described based on the example of a polygonal chain. There, the smooth mathematical line is approximated by a polygonal chain, $\mathbf{C}$.
$\mathbf{C}$ is a curve defined by a sequence of points ($\mathbf{P}_0$, $\mathbf{P}_1$,..., $\mathbf{P}_n$), and these points are called vertices. 
In addition, the curve consists of segments connecting consecutive pairs of  vertices. 
In general, we can define the shape of a segment through the \textit{shape function} which
allows to have not only straight line segments but also spline curves of different order.

\section{The Dislocation Ontology}
\label{sec4/dislocation-ontology}

\begin{figure*}[htb]
\centering
\includegraphics[width=0.9\textwidth]{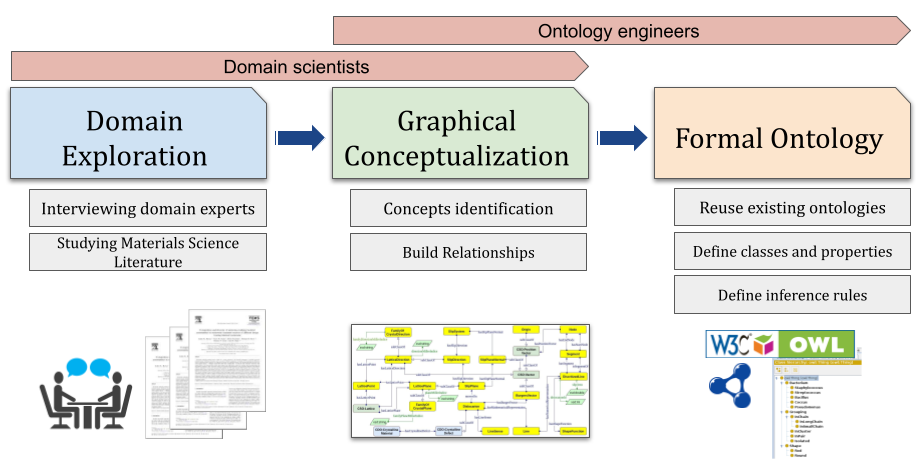}
\caption{The workflow of the dislocation ontology development, illustrating the main phases, subprocesses, and roles involved in the whole process.}
\label{fig:workflow}
\end{figure*}

The dislocation ontology, DISO, is developed using several well-known ontology development methodologies, such as \cite{suarez2015neon}. 
The process is iterative, starting with an initial version and continuously revising and refining the evolving ontology.
The development process is outlined in \autoref{fig:workflow}, which includes the main phases, their sub-tasks, and the roles involved.

\subsection{Metadata}
\label{sec:overview}


It is essential to provide a systematic and comprehensive description of the ontology, also known as ontology metadata, thus supporting its reusability and findability \cite{Simperl2011-Ontology-metadata}. 
When ontology metadata is missing, several potential issues can occur. These include reduced accessibility for potential users, decreased reusability, and ontologies not being recognized as relevant for specific use cases.
Accordingly, several DCMI Metadata Terms\footnote{{\url{https://www.dublincore.org/specifications/dublin-core/dcmi-terms/}}} have been added to the ontology, involving \texttt{terms:contributor},  \texttt{terms:created},  \texttt{terms:title}, and \texttt{vann:preferredNamespacePrefix}. 

\subsection{Reuse of existing models}
\label{sec:reuse}
 When developing an ontology, one of the first steps is to utilize or reuse terms (i.e., classes or properties) from existing ontologies that describe the same domain or subject matter. 
 Deciding which ontologies are appropriate for reuse is a challenging task for ontology engineers. 
Ontology reuse involves several activities, including merging, extending, specializing, or adapting other ontologies.
In DISO, we reuse concepts from two related ontologies in the MSE domain: the \textit{Crystal Structure Ontology}\footnote{{\url{https://purls.helmholtz-metadaten.de/disos/cso}}} (CSO) and the \textit{Crystalline Defect Ontology\footnote{{\url{https://purls.helmholtz-metadaten.de/disos/cdo}}}} (CDO). 
CSO describes crystallographic data related to dislocations, while CDO links physical material entities to crystal structures and different defect types within a crystal, such as point defect, dislocation, and planar defect.

\begin{sloppypar}
In CDO, the \texttt{EMMO:Crystal} class (from the EMMO\footnote{\url{https://github.com/emmo-repo/domain-crystallography}} ontology) is reused to describe the physical entity of crystalline materials. 
The \texttt{CDO:CrystallineMaterial} class is defined as a subclass of \texttt{EMMO:Crystal} which is used to represent crystalline materials. 
\end{sloppypar}

\begin{sloppypar}
In CSO, several MDO~\cite{li2020} classes are reused to describe the crystal coordinate system, the motif in a crystal structure, point groups, and space groups. 
Furthermore, the CSO defines the unit quantity of a property by reusing several classes from QUDT (Quantities, Units, Dimensions and Data Types Ontologies)~\cite{hodgson2014qudt}.
Overall, the semantic data value of the developed ontology increases as more ontologies are included, making the reuse of terms from other ontologies a worthwhile undertaking~\cite{seo2019}.
\end{sloppypar}


\subsection{Classes}
\label{sec4/main-classes}
Our ontology classes are separated into two groups: (i) those imported from existing ontologies (as explained in \autoref{sec:reuse}) and (ii) newly created classes that are not already defined in any existing ontologies. 

\begin{sloppypar}
\emph{Imported classes}.
DISO reuses several classes from  CSO: \texttt{CSO:Lattice} represents the periodic arrangement of one or more atoms, and \texttt{CSO:Vector} represents quantities with both magnitude and direction. 
Additionally, DISO reuses classes from  CDO, including \texttt{CDO:CrystallographicDefect}, which represents lattice irregularity or lattice defects.
\end{sloppypar}

\emph{Newly defined classes}. 
For new classes, we focus on specific classes of crystalline materials and line defects, including 
1) \texttt{Dislocation}, the focal class in the DISO which represents a linear or one-dimensional defect that causes some atoms to be displaced,
2) \texttt{SlipPlane}, which models the lattice plane to which the dislocation is constrained to move in, 
3) \texttt{SlipDirection}, which models the lattice direction where the slip occurs in the crystalline materials, 
4) \texttt{LatticePlane}, which represents the lattice plane where it forms an infinitely stretched plane that cuts through the lattice points, 
5) \texttt{LatticeDirection}, which models the direction inside the lattice that connects two lattice points, and 
6) \texttt{DiscretizedLine}, which provides a numerical representation of the dislocation line as a mathematical line, such as an oriented curve, that is discretized into several segments.

 \begin{figure*}[tb]
 \centering
     \centering
     \includegraphics[width=0.9\linewidth]{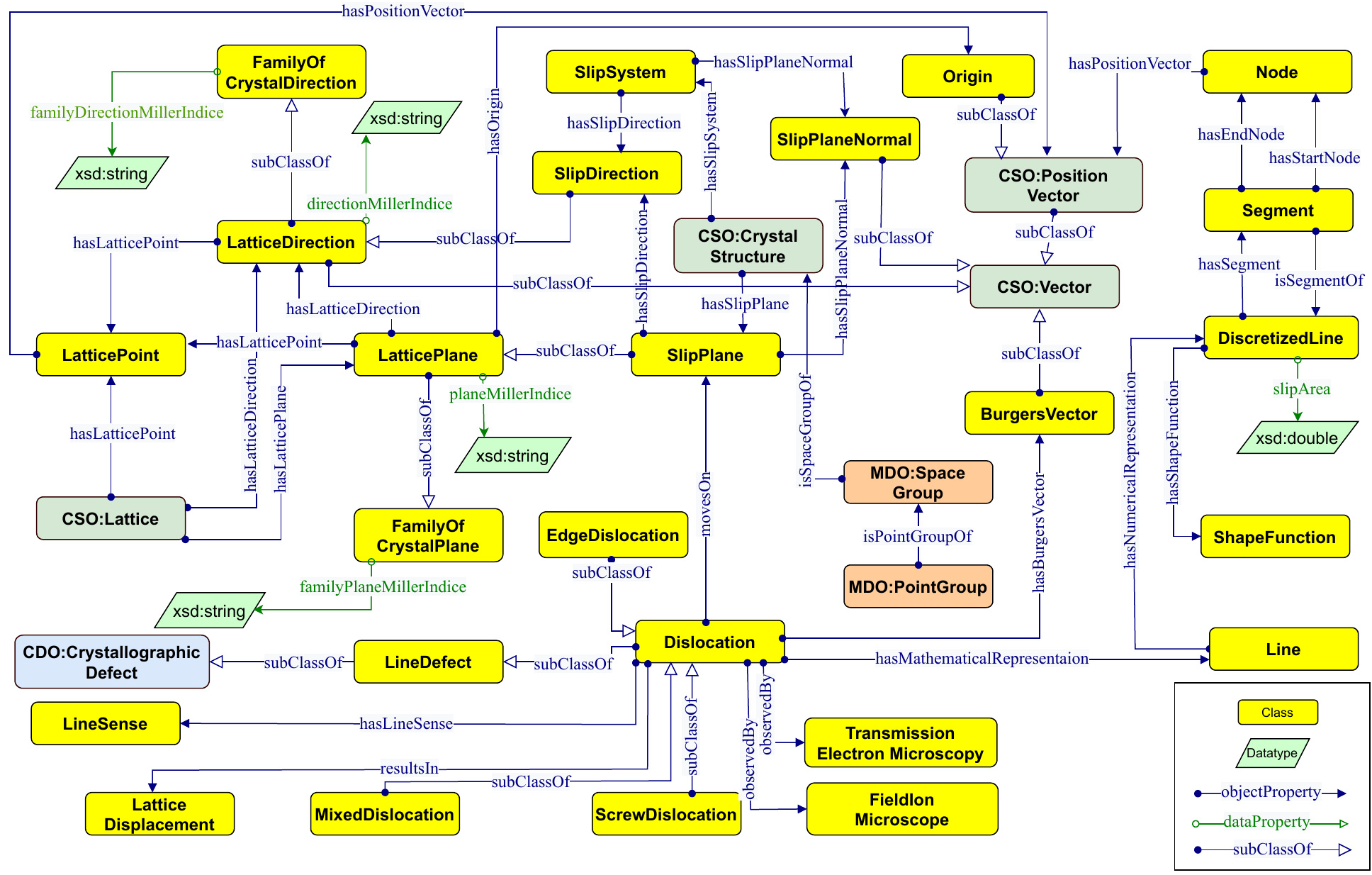}
     \caption{Core concepts and interconnected relationships in the DISO ontology. Arrows with open arrowheads denote \texttt{rdfs:subClassOf} properties between classes. Regular arrows represent \textit{rdfs:domain} and \textit{rdfs:range} restrictions on properties and coloured boxes represent classes belonging to different ontologies, e.g., yellow boxes represent DISO's classes.}
     \label{fig:mainConcepts}
 \end{figure*}
  
\subsection{Properties}

Similarly, both data and object properties in DISO are divided into two categories which are newly defined properties and reused ones.

\emph{Newly defined properties}.
Object properties constitute the relationship between various concepts in the ontology. For instance, the relationship between \texttt{TransmissionElectronMicroscopy} and \texttt{Dislocation} classes can be represented through the \texttt{observedBy} object property. 
Similarly, the \texttt{hasLineSense} object property represents the relationship between \texttt{Dislocation} and \texttt{LineSense}. 
Additionally, a number of data properties, including \texttt{directionMillerIndice} and \texttt{planeMillerIndice} are defined, which typically provide a relation to attaching an entity instance to some literal datatype value, such as a string or a date.

\emph{Reused properties}.
Several properties from the reused ontologies have been used, e.g., \texttt{cso:hasPositionVector}, \texttt{cdo:hasCrystallographicDefect}, \texttt{mdo:hasComposition}, and \texttt{emmo:hasProperty} from the CSO ontology, CDO ontology, MDO ontology, and EMMO ontology, respectively.   
Moreover, we reused several data properties from DCterms for adding ontology metadata (see \autoref{sec:overview}).
After defining new properties and identifying reused ones, the domain and range for each property using \texttt{rdfs:domain} and \texttt{rdfs:range} are defined, respectively. 
For instance, the domain of the data property \texttt{diso:planeMillerIndice} is \texttt{diso:LatticePlane} and the range is \texttt{xsd:string}.
while the domain of the object property \texttt{diso:hasLatticePoint} is \texttt{diso:LatticePlane} and the range is \texttt{diso:LatticePoint}.
\begin{figure*}[htbp]
 \centering
    	\includegraphics[width=0.95\textwidth]{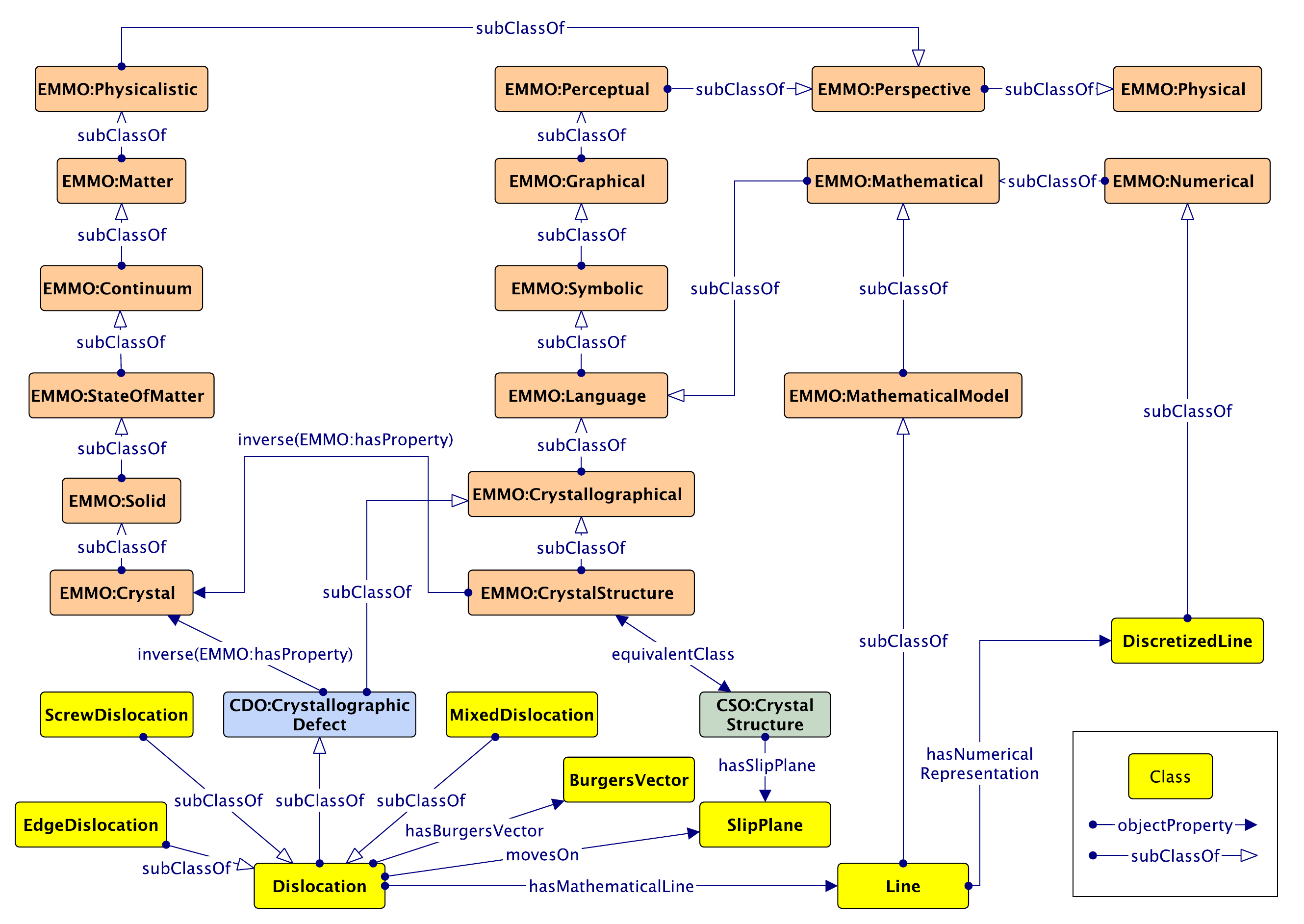}
    	\caption[]{DISO alignment with EMMO. The starting point to align the DISO with the EMMO by importing the domain ontology crystallography developed by EMMO.}
    	\label{fig/diso-emmo}
\end{figure*}

\emph{Restricting properties}. 
In the DISO, several classes use property restrictions, e.g., value constraints. 
For example, the \texttt{resultsIn} property which connects \texttt{Dislocation} and \texttt{LatticeDisplacement} is restricted by a value constraint of \texttt{owl:someValuesFrom} representing the fact that every dislocation individual results in \textit{some} or at least one lattice displacement individual(s).
The \texttt{hasLineSense} property which connects \texttt{Dislocation} and \texttt{LineSense} is restricted by a value constraint of \texttt{owl:allValuesFrom} representing that every dislocation individual can \textit{only} have a line sense individual. 

\subsection{Reasoning}
DISO's inference capability is increased through the use of several property characteristics, such as  functional relations, transitivity, and the inverse property~\cite{fathalla2023upper}. 
\texttt{hasMathematicalRepresentation} is a functional property because it means that a dislocation can be represented by exactly one mathematical line, i.e., it can not have any different mathematical representation than that.
The transitive property can be demonstrated through the \texttt{hasRepresentation} relationship. 
This relationship refers to the connection between a dislocation and its representation. For instance, if a dislocation has a line representation, and this line has a discretized line representation, it can be inferred that the dislocation also has a discretized line representation.
In order to enable bidirectional navigation between two classes in the ontology, inverse properties are established for each corresponding property.
For instance, the \texttt{isSegmentOf} property is the inverse property of \texttt{hasSegment}.
This means that if a discretized line A\textit{ has a segment} B, then B \textit{is a segment of} A.


\section{Ontology Alignment}
\label{sec5/ontology-alignment}

Ontology alignment is the process of identifying relations between entities among different ontologies in order to  establish connections between them \cite{ehrig2006ontology}. 
These entities include classes, properties, and individuals. 
For successful ontology alignment, it is crucial to identify similarities between source and target ontologies. 
The analysis entails examining concepts that overlap but may have different names (i.e. synonyms) or types in the ontologies~\cite{noy2000algorithm}.

This section will cover the extension of DISO, which involves aligning two ontologies, namely EMMO and MDO.
This alignment plays a crucial role in allowing DISO to annotate the DDD data and transform it into linked data while also facilitating knowledge graph generation.

\begin{figure*}[tb]
 \centering
    	\includegraphics[width=\textwidth]{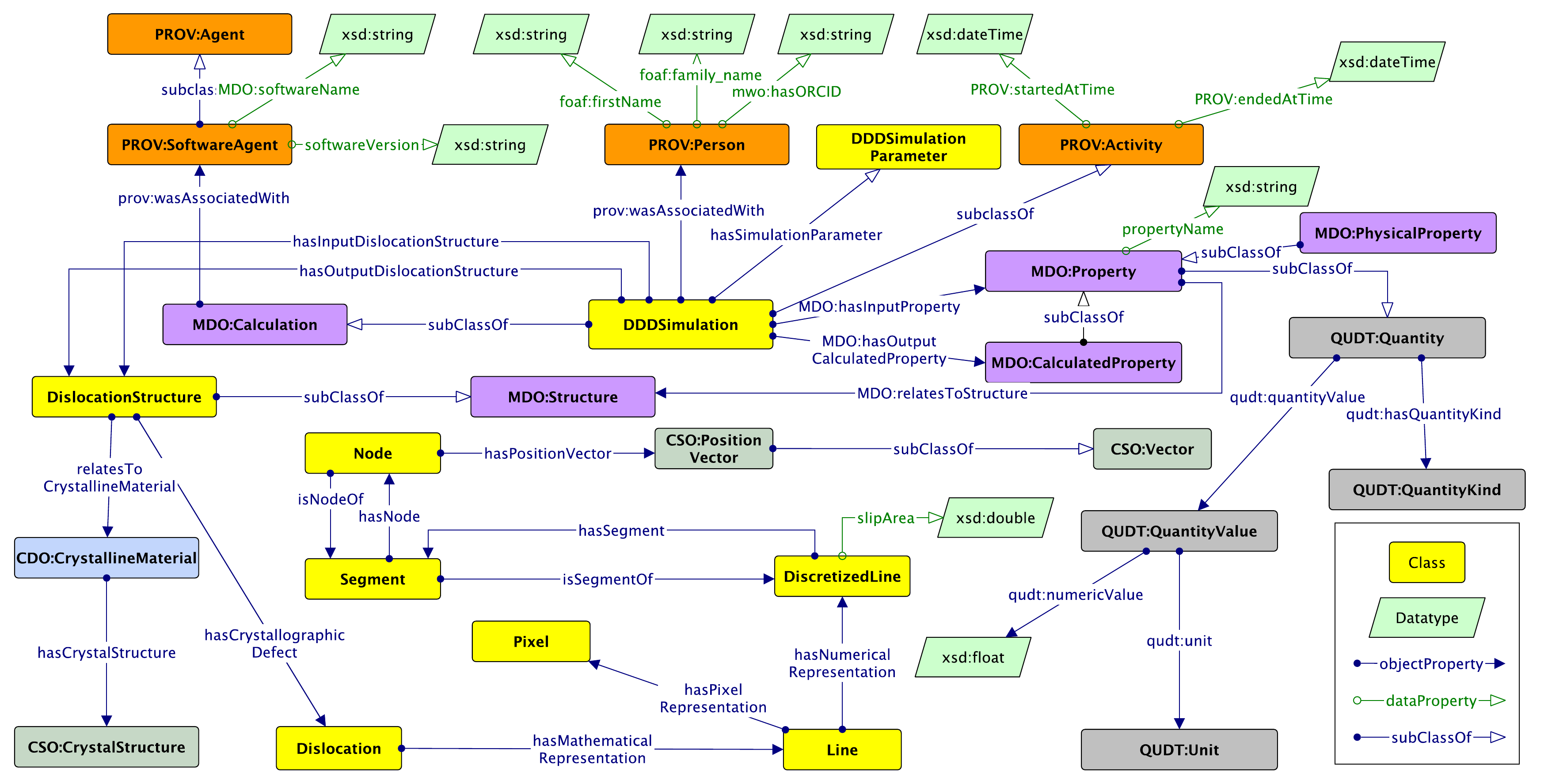}
    	\caption[]{DISO alignment with the MDO Core and Provenance module.}
    	\label{fig/diso-mdo-core}
\end{figure*}

\subsection{Alignment with EMMO}

EMMO is a continuous initiative aimed at establishing semantic standards that can be implemented at the highest level of abstraction.
This makes it possible for all potential domain ontologies, especially in the MSE field, to be integrated and to work together seamlessly.
Currently, EMMO consists of two modules: a top-level and a mid-level module. 
The former includes the fundamental axioms that constitute the philosophical foundation of the EMMO, while the latter consists of a set of perspectives to develop more specialized domain ontologies. 
These two ontologies serve as the basis for building further domain and applications ontologies, e.g., the application of EMMO in the domain of mechanical testing~\cite{morgado2020}.

The starting point to align the DISO with EMMO is by aligning with the Crystallography Domain Ontology\footnote{\url{https://github.com/emmo-repo/domain-crystallography}}, a domain ontology based on EMMO and the CIF core dictionary\footnote{\url{https://www.iucr.org/resources/cif/dictionaries/cif_core}}.
As shown in \autoref{fig/diso-emmo}, \texttt{CDO:CrystallographicDefect} subsumes \texttt{Dislocation}, while also being a subclass of \texttt{EMMO:Crystallographical} class.
Similarly, \texttt{EMMO:CrystalStructure}, an equivalent class to \texttt{CSO:CrystalStructure}, is also a subclass of \texttt{EMMO:Crystallographical}. 
Overall, \texttt{EMMO:Crystallographical} is a class that ideally represents the physical concepts associated with crystalline materials.

As we mentioned in \autoref{sec3/domain-description}, on the mesoscale, a dislocation is represented by a mathematical line, which can be further idealized as a pixel or discretized line depending on the application (e.g., microscopy or simulation).
To align the dislocation mathematical line concept with an EMMO class, \texttt{EMMO:MathematicalModel} subsumes \texttt{Line} and the discretized representation of the mathematical dislocation line, \texttt{DiscretizedLine}, is subsumed \texttt{EMMO:Numerical}.

\subsection{Alignment with MDO}

The MDO is a domain ontology that defines concepts and relations to cover the knowledge of materials design, especially in the ab-initio calculation.
MDO consists of several modules, a \textit{Core}, the \textit{Provenance} module, and two domain-specific modules: \textit{Structure} and \textit{Calculation}. 

To align the DISO with the MDO, we reused several classes in the MDO Core module.
The MDO Core module describes the structure or the virtual specimen of interest via \texttt{MDO:Structure} class. 
As shown in \autoref{fig/diso-mdo-core}, we defined a \texttt{DislocationStructure} class as a subclass of \texttt{MDO:Structure}.
This class describes a dislocation (micro)structure, which is a virtual specimen used by a DDD simulation to study the mechanical properties of a  crystalline material.  
Furthermore, \texttt{DislocationStructure} as an idealized representation relates to a physical concept called \texttt{CDO:CrystallineMaterial}.

 \begin{figure*}[tb]
 \centering
     \includegraphics[width=\linewidth]{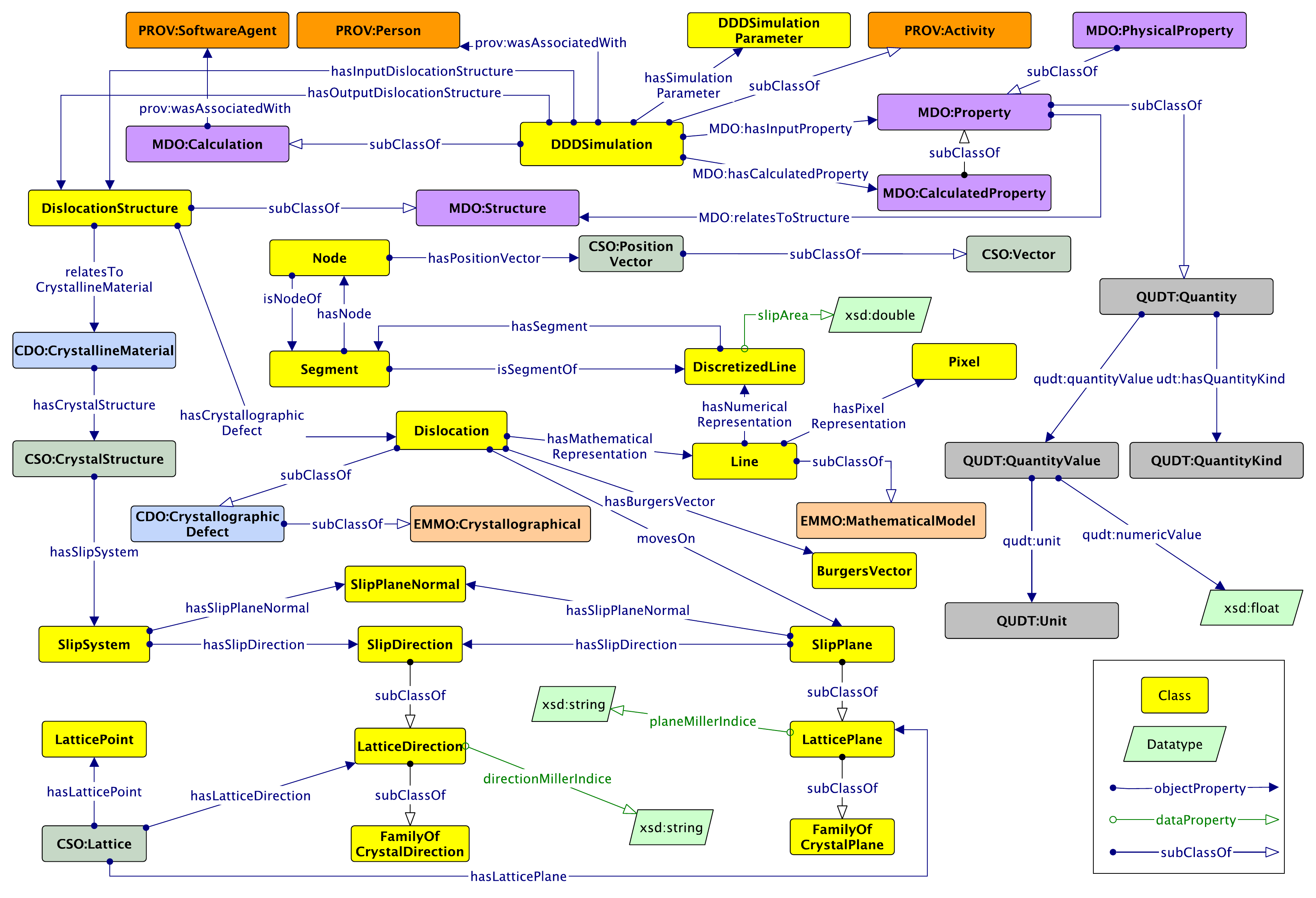}
     \caption{The core concepts in DISO ontology after the alignment with MDO and EMMO. Arrows with open arrowheads denote \texttt{rdfs:subClassOf} properties between classes, while regular arrows represent the relationships between them. Classes that belong to the same ontology share the same color.}
     \label{fig:diso-extend}
 \end{figure*}

In the MDO Core module, an instance of the \texttt{MDO:Structure} class is used as a virtual specimen input or output for a simulation.
Here, the simulation concept is represented as the \texttt{MDO:Calculation} class. 
We subsumed the \texttt{MDO:Calculation} class to define the \texttt{DDDSimulation} class, which is a class to describe the DDD simulation. 
Thus, the \texttt{DDDSimulation} can have \texttt{DislocationStructure} as an input or an output. 
Moreover, the \texttt{DDDSimulation} has an input and output relationship with \texttt{MDO:Property} to run a calculation.
In addition, the \texttt{DDDSimulation} is related to the  \texttt{DDDSimulationParameter}, a simulation parameter concept configuring the DDD simulation, e.g., the activation parameter for cross-slip, junction formation, and external load.
\begin{figure*}[tb]
 \centering
    	\includegraphics[width=0.5\linewidth]{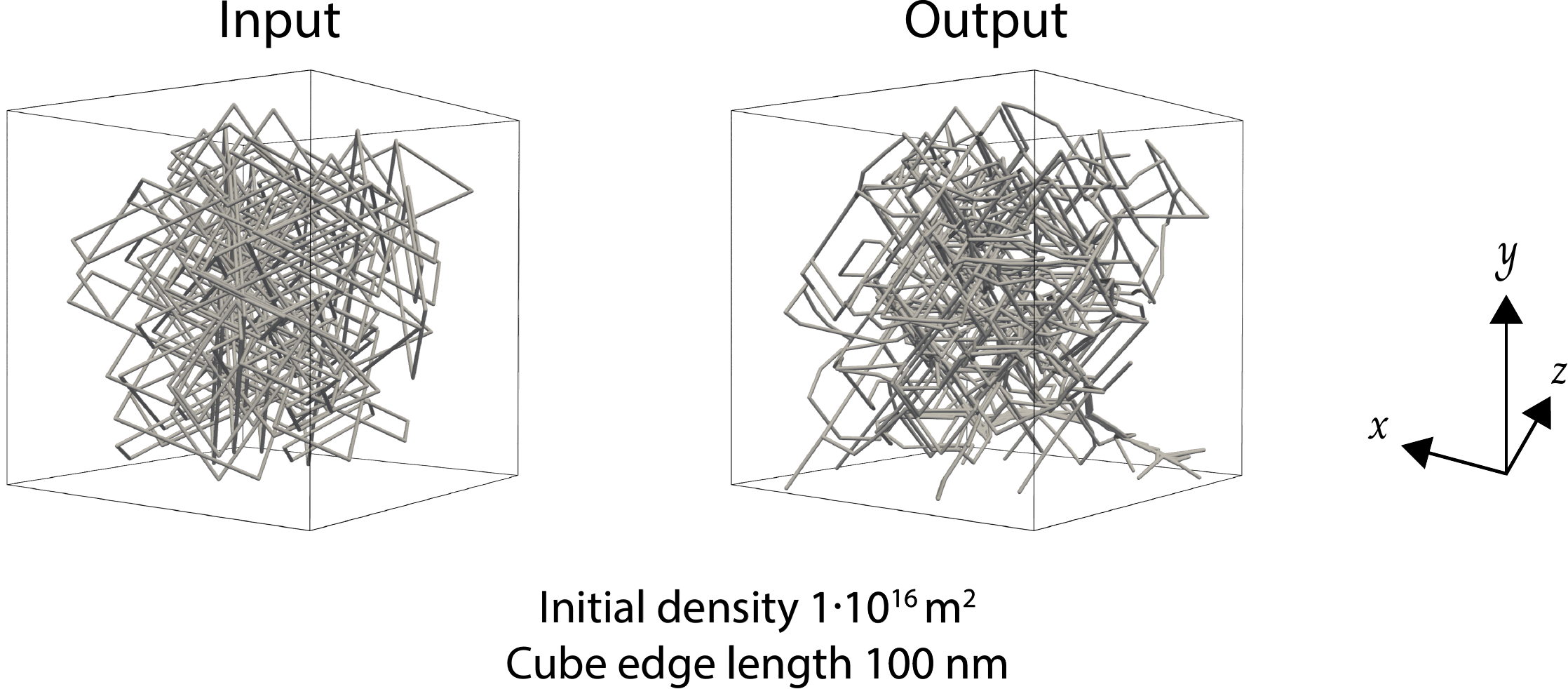}
    	\caption[]{Sample of dislocation microstructures used as input for simulation as well as yielded by the simulation as output.}
    	\label{fig/cu_1}
\end{figure*}
To preserve the provenance information of a DDD simulation, we reused several classes from the MDO Provenance module and the PROV  ontology~\cite{lebo2013prov}. 
Running a DDD simulation requires specific software to solve materials science problems.
It is quite helpful to store information about the software used and its version, as this can help scientists reuse data through post-processing methods specific to DDD software.
In this regard, \texttt{DDDSimulation} has a relationship with \texttt{PROV:SoftwareAgent}, which has two data properties: \texttt{softwareVersion} and \texttt{MDO:softwareName}. 
Furthermore, to preserve the provenance information related to when a DDD simulation starts and ends, \texttt{PROV:Activity} subsumed the \texttt{DDDSimulation} and inherited two data properties: \texttt{PROV:startedAtTime} and \texttt{PROV:endedAtTime}.
Apart from that, we reused \texttt{PROV:Person} to annotate the person running or responsible for the simulation. 
It has three data properties to define a person: \texttt{FOAF:firstName}, \texttt{FOAF:family\_name}, and \texttt{MWO:hasORCID}. 
The latter is a data property that we reused from the MatWerk Ontology (MWO)\footnote{{\url{hhttp://purls.helmholtz-metadaten.de/mwo/}}}.

To summarize, core concepts and interconnected relationships in DISO after the alignment can be seen in \autoref{fig:diso-extend}. 
The advantages of ontology alignment for DISO are promoting knowledge transfer from other ontologies when describing the DDD simulation. 
Furthermore, ontology alignment fosters interoperability between ontologies in MSE-related domains. 
The objective is to assist in building a knowledge graph for the dislocation domain.

\section{The Dislocation Knowledge Graph}
\label{sec6/real-world-use-case}
In the field of materials science, researchers use a numerical method called DDD simulation to analyze the behaviour of dislocations within crystalline materials. This technique helps identify the specific characteristics of each dislocation, as well as their interaction, arrangement, and collective behaviour within the material.
The simulation observes the motion and interaction of many dislocations which ultimately creates the relationship between the microstructure, loading conditions, and the mechanical properties of a crystalline material.
For simulations of dislocations, there are various software options available such as MoDELib~\cite{Po2014_1}, ParaDiS~\cite{arsenlis2007enabling}, and microMegas~\cite{devincre2011}.
Every software has a distinct collection of metadata that organizes the inputs and outputs of the simulation.

DISO was utilized in this specific scenario to accurately annotate the information collected from various DDD simulations. The ultimate goal is to generate a comprehensive dislocation knowledge graph (DisLocKG) using this data.
The DDD data used in this work was generated through the MoDELib software and took different initial dislocation densities and specimen sizes into account. 
The cube-shaped Copper specimen, with an edge length of either 50 or 100 nanometers, was randomly filled with dipolar edge loops on all slip systems until the initial density of either $1\cdot10^{16}$ m$^{-2}$ or $5\cdot10^{16}$ m$^{-2}$ was reached. 
A sample of the generated cube-shaped Copper specimen can be seen on the left panel of the \autoref{fig/cu_1}. 
During the simulation, the dislocation microstructure was allowed to relax without any external load, meaning that internal stress and image forces solely influenced the dislocation evolution.
The simulation resulted in the relaxed dislocation microstructure shown on the right-hand side of \autoref{fig/cu_1}.
An important aspect for a materials scientist is that some simulations do not have cross-slip or junction formation. This has significant implications when it comes to analyzing simulation results. E.g., \citet{demirci} investigated the influence of cross-slip on the evolution of dislocation structures and therefore could benefit from this information.
Such relaxation calculations are also important for creating a realistic microstructure.
For example, ~\citet{motz2009initial} investigated how the relaxed dislocation microstructure influences the plasticity in subsequent tensile test simulations. Furthermore, several authors~\cite{song2021data,fan2021strain,steinberger2020,stricker2018dislocation} conducted  machine learning and data mining studies utilizing the dislocation relaxed microstructure to classify the structure and express the strain energy density of a dislocation microstructure, respectively. This explains why there is a strong need for a detailed and formal representation of such simulations -- here, contained in the class of \texttt{DDDSimulation}. 

\begin{figure*}[tb]
 \centering
    	\includegraphics[width=.9\textwidth]{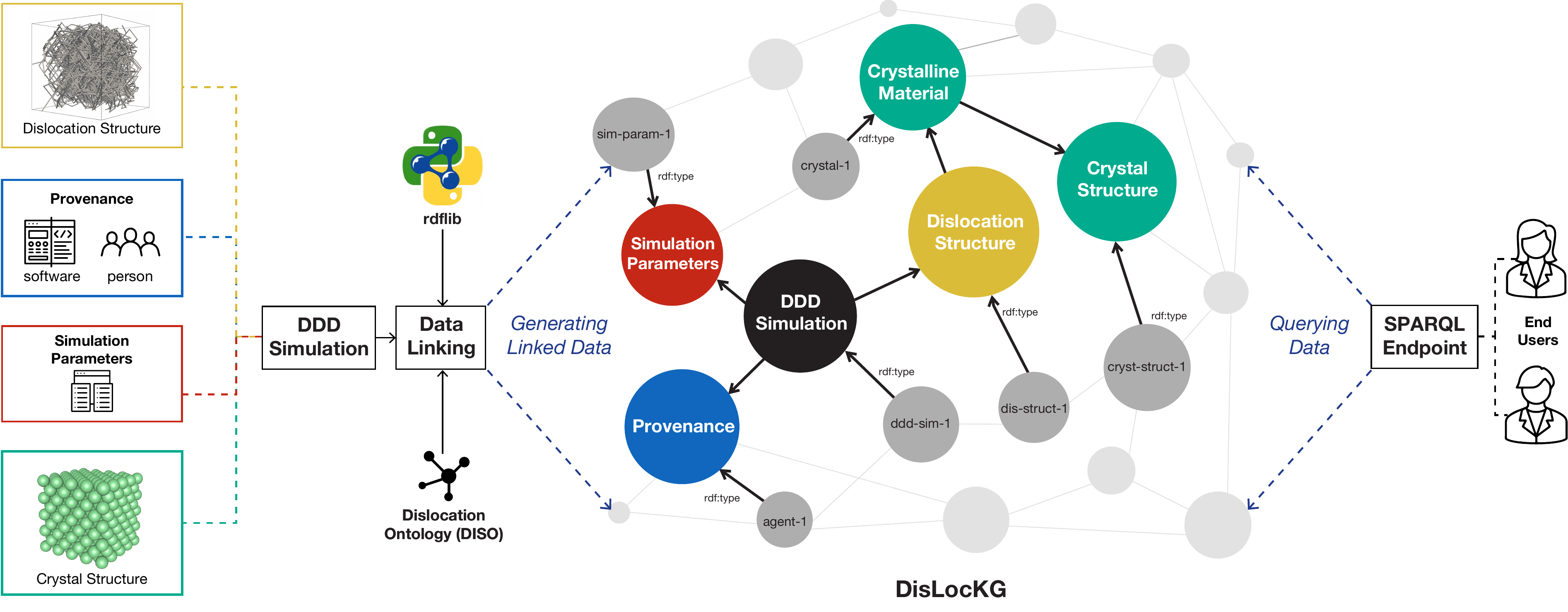}
    	\caption[]{DDD simulation data as linked data. Colored rectangles on the left depict data types in the DDD simulation: dislocation structure, provenance, simulation parameters, and crystal structure data. The data subsequently is linked using the DISO as a reference ontology. The rdflib Python module supports data linking and generates the \textit{DisLocKG}. Via the SPARQL Endpoint, end users can query the data to retrieve the information in the DisLocKG.}
    	\label{fig/diso-kg}
\end{figure*}

For our example, we have collected a total of 25 data points -- where each data point is one DDD simulation consisting of initial and final microstructure. Each of those was annotated with DISO.
Any data point gives information about the simulation details, such as parameters used for a simulation, initial dislocation microstructure used as input, and the resulting dislocation microstructure produced by the simulation. 
Additionally, each dislocation microstructure includes information about the crystal structure, Bravais lattice, dislocation, slip plane, Burgers vector, and numerical representation of dislocation. 
The results of the simulations are stored and parsed in the HDF5\footnote{\url{https://www.hdfgroup.org/solutions/hdf5/}} format. 
Subsequently, we utilize our in-house Python scripts (using the rdflib 6.0~\cite{rdflib} Python library) to create a knowledge graph called \textit{DisLocKG} from this data using DISO as a reference ontology (cf. \autoref{fig/diso-kg}). 
The DisLocKG is a semantic network that holds information about dislocations in crystalline materials.
Note that, DisLocKG also stores the provenance information related to the data, particularly the creator data, software, and software version used to generate the data. 
In total, we have generated a number of $\sim2.2$M triples that are stored as RDF files which are available via its persistent identifier\footnote{\url{https://purls.helmholtz-metadaten.de/dislockg}}.


\begin{figure*}[tb]
 \centering
    	\includegraphics[width=0.75\textwidth]{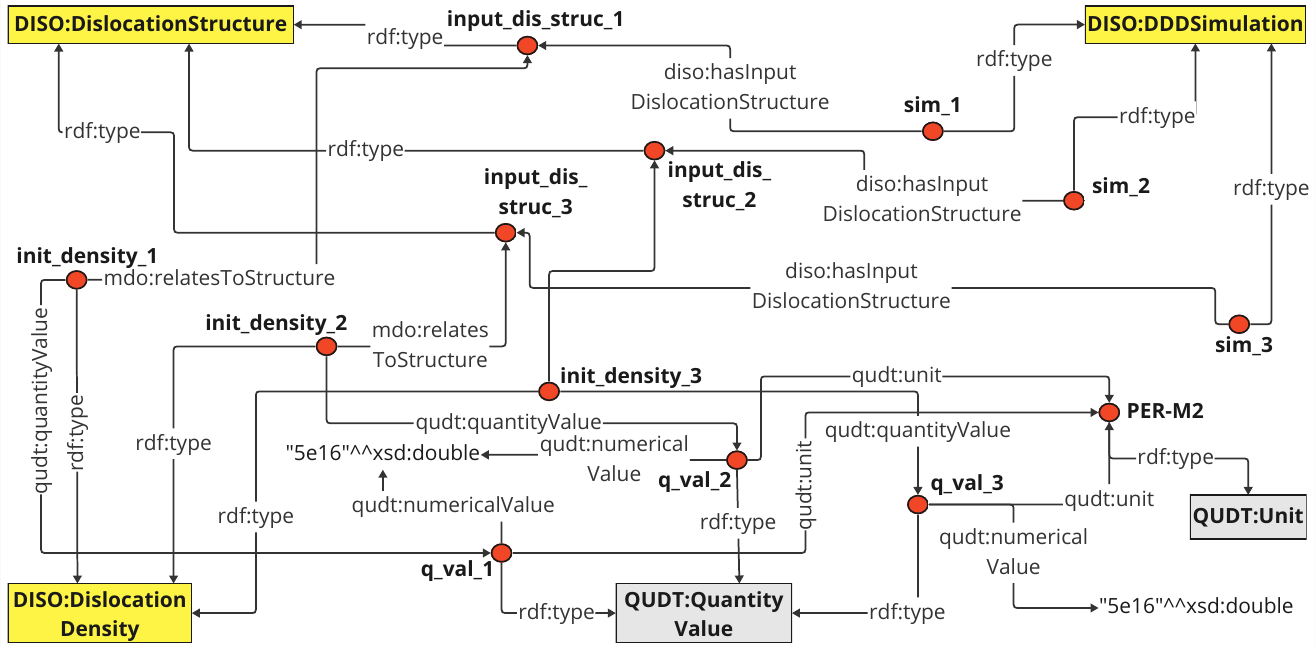}
    	\caption[]{A visual representation of CQ3 results. Colored boxes represent classes and the red dot represents an individual belonging to that class. 
    	Each individual is defined by a directed arrow having the \texttt{rdf:type} relationship to the respective class and connected to other individuals by object properties.}
    	\label{fig/vis-cq3}
\end{figure*}

Publishing DDD data as linked data has several benefits~\cite{fathalla2019eventskg}, including
1) establishing links between dislocation-related datasets, enabling machines to understand and discover new information, 
2) supporting semantic querying via the SPARQL query language,
3) supporting data enrichment, where machines can infer implicit knowledge that does not exist, and 
4) promoting semantic validation of the data, ensuring consistency and accuracy.


We have listed some competency questions in \autoref{tab:CQs} to give an idea of the vast information available in DisLocKG.
For instance, CQ1 can retrieve the history and origin information of DDD data generated by the MoDELib software, and CQ2 and CQ3 can retrieve information on the specimen geometry and the initial dislocation of each dislocation simulation. 
These CQs are important if one wants to query a dislocation simulation to be reused for the processing step if they need a specific density of a dislocation structure and information concerning the geometry.
CQ4 retrieves the input parameters to run the simulation, while CQ5 queries all dislocation structures generated by the relaxation calculation.
The SPARQL query corresponding to CQ3 is shown in \autoref{lst:SPARQL1}, and the complete set of the competency questions and the corresponding SPARQL queries can also be found in the DISO GitHub repository.
\autoref{fig/vis-cq3} visualizes the results of CQ3, which contains three individuals (shown as the red markers) of the DDD simulation class. 
Each of the DDD simulation individuals has a relationship with dislocation structure individuals. 
Moreover, the dislocation density data relates to the dislocation structure individual.

\begin{table*}
\centering
\caption{A sample of competency questions for DisLocKG.}
    \begin{tabularx}{\textwidth}{cX} 
    \hline
    \textbf{No.} & \textbf{ Question}  \\
    \hline
    CQ1 & Provide detailed information on the dislocation structures simulated using the MODELIB software, including the software version and creator associated with these simulations. \\
    CQ2 & Which dislocation structures possess a specimen shape resembling a cube with an edge length greater than 30 nanometers? \\
    CQ3 & List all DDD simulations that have an initial density of dislocation $= 5e16$ m$^{-2}$\\
    CQ4 & List all DDD simulations that do not activate the cross slip formation and junction formation\\
    CQ5 & What are dislocation structures generated by the relaxation calculation? List also the initial density of a dislocation structure used for a relaxation calculation, simulation parameters: cross-slip activation, junction formation activation, and external load activation.\\
    \hline
    \end{tabularx}
\label{tab:CQs}
\end{table*}
\begin{figure*}
\begin{lstlisting}[language=sparql,keepspaces=true,caption={SPARQL query corresponding to CQ3.},
label={lst:SPARQL1}, frame=single]
PREFIX diso:<https://purls.helmholtz-metadaten.de/disos/diso#>
PREFIX qudt:<http://qudt.org/schema/qudt/>
PREFIX xsd:<http://www.w3.org/2001/XMLSchema#>
PREFIX mdo_core:<https://w3id.org/mdo/core/>
SELECT ?ddd_simulation ?initial_density ?unit 
WHERE{
?ddd_simulation       diso:hasInputDislocationStructure   ?input_dislocstructure.
?initial_density_qu   mdo_core:relatesToStructure         ?input_dislocstructure; 
                      qudt:quantityValue                  ?density_qv.
?density_qv           qudt:numericalValue                 ?initial_density;     
                      qudt:unit                           ?unit.
FILTER(?initial_density = "5e16"^^xsd:double) 
}
\end{lstlisting}
\end{figure*}

\section{Evaluation}
\label{sec7/evaluation}
Employing predefined metrics that evaluate an ontology's richness through criteria-based assessment is one way of evaluating its quality \cite{seo2019}. 
In this section, we evaluate the adapted version of DISO using the OntoQA~\cite{tartir200} evaluation model.
\begin{table*}[htbp]
    \centering
    \begin{tabular}{ccccc|ccc} 
 \hline\hline
 Ontology & C & SC & AT & P & RR & AR & IR \\ 
 \hline
 MDO & 37 & 49 & 32 & 32 & 0.40 & 0.86 & 1.32\\
 CSO & 30 & 49 & 19 & 25 & 0.34 & 0.63 & 1.63\\
 DISO v1.0 & 33 & 62 & 12 & 33 & 0.35 & 0.32 & 1.63\\
 \hline
 \textbf{DISO v1.1 }& \textbf{70} & \textbf{116} & \textbf{47} & \textbf{80} & \textbf{0.41} & \textbf{0.67} & \textbf{1.66} \\
 \hline\hline\\
\end{tabular}
    \caption{Evaluation of DISO compared to DISO v1.0, MDO and CSO using the OntoQA model. C is the number of classes, SC is the number of sub-classes, AT is the number of attributes, and P denotes the number of relationships.}
\label{tab:ontoqa}
\end{table*}
This model can assess an ontology based on two dimensions: Schema and Instances. 
Here, we focus on the schema evaluation which evaluates the quality of the ontology's design.
We determine the effectiveness of the ontology and its ability to represent rich knowledge using the following metrics:
\begin{itemize}
    \item \emph{Relationship richness ({RR})} shows the diversity of relations and placement of relations in the ontology (\autoref{eq:RR}). 
    
    \begin{equation}
        RR = \frac{|P|}{|SC|+|P|}
        \label{eq:RR}
    \end{equation} 
    where \textit{P} is the number of relationships and \textit{SC} is the number of sub-classes. The more relations an ontology owns, the richer it is (\emph{is-a} relations are not considered). 
    \item \emph{Attribute richness ({AR})} shows that the more attributes are defined the more knowledge the ontology delivers (\autoref{eq:AR}).   
    
    \begin{equation}
        AR = \frac{|AT|}{|C|}
        \label{eq:AR}
    \end{equation} where \textit{AT} is the number of attributes for all classes and \textit{C} is the number of classes.
    
    \item \emph{Inheritance richness (\textit{IR})} describes the distribution of information across different levels of the ontology inheritance tree. IR indicates how knowledge is classified into different classes and subclasses in an ontology. (\autoref{eq:IR}). 
    
    \begin{equation}
        IR = \frac{|SC|}{|C|}
        \label{eq:IR}
    \end{equation} 
\end{itemize}

In \autoref{tab:ontoqa}, we compare the evaluation outcomes of DISO with MDO~\cite{li2020}, CSO\footnote{\url{https://purls.helmholtz-metadaten.de/disos/cso}} and the previous version of DISO.
DISO has the most significant value of \textit{RR}, which implies that it has a greater relation diversity. 
Moreover, DISO has the highest \textit{IR} value, representing a more comprehensive knowledge range than MDO, CSO, and the previous version of DISO. 
The \textit{AR} value of DISO is lower than that of MDO and higher than CSO and the previous version of DISO.
To conclude, DISO possesses the most extensive knowledge representation and diversity in terms of relationships, achieving the highest IR and RR, respectively.
Moreover, the adapted version of DISO surpasses its predecessor in all evaluation metrics.

\section{Conclusion and Outlook}
\label{sec8/conclusion}
This paper showcases how semantic web technologies can be utilized to transform unstructured DDD data into well-organized and structured data.
Furthermore, we extended the dislocation ontology by aligning it with commonly used materials science ontologies (i.e.,  EMMO and MDO core) to be able to model simulation data efficiently.
Moreover, we presented a real-world use case that utilized the DISO to construct a semantic network of DDD data (i.e., linked data) called DisLocKG, where individual entities are connected, enabling semantic query and supporting intelligent tasks. 
To support querying DisLocKG, the graph has been made publicly available and steps to set up a SPARQL endpoint are described in its GitHub repository. 
The evaluation results indicate that the adapted version of DISO is the most comprehensive and diverse knowledge representation among the state-of-the-art ontologies.

In the future, we plan to improve DISO by modelling the linear elasticity theory of dislocations and extending the real-world use case, e.g., another DDD simulation software and TEM data. 
In addition, developing the DisLocKG Application Programming Interface (API) will also be a worthwhile undertaking. 
The idea is to develop several interactive features, e.g., querying, data mining, visualizing, updating, and deleting data with the DisLocKG via its API. 
DISO and DisLocKG will continue to be maintained and extended in the context of the Helmholtz Metadata Collaboration (HMC) and NFDI-MatWerk efforts of facilitating machine readability and reuse of research data.

\backmatter

\bmhead{Acknowledgments}
AI and SS acknowledge financial support from the European Research Council through the ERC Grant Agreement No. 759419 MuDiLingo ("A Multiscale Dislocation Language for Data-Driven Materials Science").
AI, SF, and SS acknowledge Helmholtz Metadata Collaboration (HMC) within the Hub Information at the Forschungszentrum Jülich (FZJ).
We are grateful to Aytekin Demirci for the MoDELib parser that we used.

\section*{Declarations}
\bmhead{Conflict of interest} The authors have no conflicts of interest to declare.








\bibliography{references} 


\begin{thebibliography}{50}
\ifx \bisbn   \undefined \def \bisbn  #1{ISBN #1}\fi
\ifx \binits  \undefined \def \binits#1{#1}\fi
\ifx \bauthor  \undefined \def \bauthor#1{#1}\fi
\ifx \batitle  \undefined \def \batitle#1{#1}\fi
\ifx \bjtitle  \undefined \def \bjtitle#1{#1}\fi
\ifx \bvolume  \undefined \def \bvolume#1{\textbf{#1}}\fi
\ifx \byear  \undefined \def \byear#1{#1}\fi
\ifx \bissue  \undefined \def \bissue#1{#1}\fi
\ifx \bfpage  \undefined \def \bfpage#1{#1}\fi
\ifx \blpage  \undefined \def \blpage #1{#1}\fi
\ifx \burl  \undefined \def \burl#1{\textsf{#1}}\fi
\ifx \doiurl  \undefined \def \doiurl#1{\url{https://doi.org/#1}}\fi
\ifx \betal  \undefined \def \betal{\textit{et al.}}\fi
\ifx \binstitute  \undefined \def \binstitute#1{#1}\fi
\ifx \binstitutionaled  \undefined \def \binstitutionaled#1{#1}\fi
\ifx \bctitle  \undefined \def \bctitle#1{#1}\fi
\ifx \beditor  \undefined \def \beditor#1{#1}\fi
\ifx \bpublisher  \undefined \def \bpublisher#1{#1}\fi
\ifx \bbtitle  \undefined \def \bbtitle#1{#1}\fi
\ifx \bedition  \undefined \def \bedition#1{#1}\fi
\ifx \bseriesno  \undefined \def \bseriesno#1{#1}\fi
\ifx \blocation  \undefined \def \blocation#1{#1}\fi
\ifx \bsertitle  \undefined \def \bsertitle#1{#1}\fi
\ifx \bsnm \undefined \def \bsnm#1{#1}\fi
\ifx \bsuffix \undefined \def \bsuffix#1{#1}\fi
\ifx \bparticle \undefined \def \bparticle#1{#1}\fi
\ifx \barticle \undefined \def \barticle#1{#1}\fi
\bibcommenthead
\ifx \bconfdate \undefined \def \bconfdate #1{#1}\fi
\ifx \botherref \undefined \def \botherref #1{#1}\fi
\ifx \url \undefined \def \url#1{\textsf{#1}}\fi
\ifx \bchapter \undefined \def \bchapter#1{#1}\fi
\ifx \bbook \undefined \def \bbook#1{#1}\fi
\ifx \bcomment \undefined \def \bcomment#1{#1}\fi
\ifx \oauthor \undefined \def \oauthor#1{#1}\fi
\ifx \citeauthoryear \undefined \def \citeauthoryear#1{#1}\fi
\ifx \endbibitem  \undefined \def \endbibitem {}\fi
\ifx \bconflocation  \undefined \def \bconflocation#1{#1}\fi
\ifx \arxivurl  \undefined \def \arxivurl#1{\textsf{#1}}\fi
\csname PreBibitemsHook\endcsname

\bibitem[\protect\citeauthoryear{Taylor}{1934}]{taylor1934a}
\begin{barticle}
\bauthor{\bsnm{Taylor}, \binits{G.I.}}:
\batitle{The mechanism of plastic deformation of crystals. part
  {I.--Theoretical}}.
\bjtitle{Proceedings of the Royal Society of London. Series A, Containing
  Papers of a Mathematical and Physical Character}
\bvolume{145}(\bissue{855}),
\bfpage{362}--\blpage{387}
(\byear{1934})
\end{barticle}
\endbibitem

\bibitem[\protect\citeauthoryear{Polanyi}{1934}]{polanyi1934}
\begin{barticle}
\bauthor{\bsnm{Polanyi}, \binits{M.}}:
\batitle{{{\"U}ber eine Art Gitterst{\"o}rung, die einen Kristall plastisch
  machen k{\"o}nnte}}.
\bjtitle{Zeitschrift f{\"u}r Physik}
\bvolume{89}(\bissue{9-10}),
\bfpage{660}--\blpage{664}
(\byear{1934})
\end{barticle}
\endbibitem

\bibitem[\protect\citeauthoryear{Callister and Rethwisch}{2018}]{callister2018}
\begin{bbook}
\bauthor{\bsnm{Callister}, \binits{W.D.}},
\bauthor{\bsnm{Rethwisch}, \binits{D.G.}}:
\bbtitle{Materials Science and Engineering: an Introduction}
vol. \bseriesno{9}.
\bpublisher{Wiley New York}, \blocation{???}
(\byear{2018})
\end{bbook}
\endbibitem

\bibitem[\protect\citeauthoryear{Murakumo et~al.}{2004}]{murakumo2004creep}
\begin{barticle}
\bauthor{\bsnm{Murakumo}, \binits{T.}},
\bauthor{\bsnm{Kobayashi}, \binits{T.}},
\bauthor{\bsnm{Koizumi}, \binits{Y.}},
\bauthor{\bsnm{Harada}, \binits{H.}}:
\batitle{Creep behaviour of ni-base single-crystal superalloys with various
  $\gamma$' volume fraction}.
\bjtitle{Acta Materialia}
\bvolume{52}(\bissue{12}),
\bfpage{3737}--\blpage{3744}
(\byear{2004})
\end{barticle}
\endbibitem

\bibitem[\protect\citeauthoryear{Govind et~al.}{2023}]{govind2023deep}
\begin{botherref}
\oauthor{\bsnm{Govind}, \binits{K.}},
\oauthor{\bsnm{Oliveros}, \binits{D.}},
\oauthor{\bsnm{Dlouhy}, \binits{A.}},
\oauthor{\bsnm{Legros}, \binits{M.}},
\oauthor{\bsnm{Sandfeld}, \binits{S.}}:
Deep learning of crystalline defects from tem images: A solution for the
  problem of" never enough training data".
arXiv preprint arXiv:2307.06322
(2023)
\end{botherref}
\endbibitem

\bibitem[\protect\citeauthoryear{Bertin and
  Zhou}{2023}]{bertin2023accelerating}
\begin{barticle}
\bauthor{\bsnm{Bertin}, \binits{N.}},
\bauthor{\bsnm{Zhou}, \binits{F.}}:
\batitle{Accelerating discrete dislocation dynamics simulations with graph
  neural networks}.
\bjtitle{Journal of Computational Physics}
\bvolume{487},
\bfpage{112180}
(\byear{2023})
\end{barticle}
\endbibitem

\bibitem[\protect\citeauthoryear{Zhang et~al.}{2022}]{zhang2022data}
\begin{barticle}
\bauthor{\bsnm{Zhang}, \binits{C.}},
\bauthor{\bsnm{Song}, \binits{H.}},
\bauthor{\bsnm{Oliveros}, \binits{D.}},
\bauthor{\bsnm{Fraczkiewicz}, \binits{A.}},
\bauthor{\bsnm{Legros}, \binits{M.}},
\bauthor{\bsnm{Sandfeld}, \binits{S.}}:
\batitle{Data-mining of in-situ tem experiments: On the dynamics of
  dislocations in cocrfemnni alloys}.
\bjtitle{Acta Materialia}
\bvolume{241},
\bfpage{118394}
(\byear{2022})
\end{barticle}
\endbibitem

\bibitem[\protect\citeauthoryear{Yang et~al.}{2020}]{yang2020learning}
\begin{barticle}
\bauthor{\bsnm{Yang}, \binits{Z.}},
\bauthor{\bsnm{Papanikolaou}, \binits{S.}},
\bauthor{\bsnm{Reid}, \binits{A.C.}},
\bauthor{\bsnm{Liao}, \binits{W.-k.}},
\bauthor{\bsnm{Choudhary}, \binits{A.N.}},
\bauthor{\bsnm{Campbell}, \binits{C.}},
\bauthor{\bsnm{Agrawal}, \binits{A.}}:
\batitle{Learning to predict crystal plasticity at the nanoscale: Deep residual
  networks and size effects in uniaxial compression discrete dislocation
  simulations}.
\bjtitle{Scientific reports}
\bvolume{10}(\bissue{1}),
\bfpage{8262}
(\byear{2020})
\end{barticle}
\endbibitem

\bibitem[\protect\citeauthoryear{Song et~al.}{2021}]{song2021data}
\begin{barticle}
\bauthor{\bsnm{Song}, \binits{H.}},
\bauthor{\bsnm{Gunkelmann}, \binits{N.}},
\bauthor{\bsnm{Po}, \binits{G.}},
\bauthor{\bsnm{Sandfeld}, \binits{S.}}:
\batitle{Data-mining of dislocation microstructures: concepts for
  coarse-graining of internal energies}.
\bjtitle{Modelling and Simulation in Materials Science and Engineering}
\bvolume{29}(\bissue{3}),
\bfpage{035005}
(\byear{2021})
\end{barticle}
\endbibitem

\bibitem[\protect\citeauthoryear{Salmenjoki
  et~al.}{2018}]{salmenjoki2018machine}
\begin{barticle}
\bauthor{\bsnm{Salmenjoki}, \binits{H.}},
\bauthor{\bsnm{Alava}, \binits{M.J.}},
\bauthor{\bsnm{Laurson}, \binits{L.}}:
\batitle{Machine learning plastic deformation of crystals}.
\bjtitle{Nature communications}
\bvolume{9}(\bissue{1}),
\bfpage{5307}
(\byear{2018})
\end{barticle}
\endbibitem

\bibitem[\protect\citeauthoryear{Prakash and
  Sandfeld}{2018}]{Prakash2018_PrM55}
\begin{barticle}
\bauthor{\bsnm{Prakash}, \binits{A.}},
\bauthor{\bsnm{Sandfeld}, \binits{S.}}:
\batitle{Chances and challenges in fusing data science with materials science}.
\bjtitle{Practical Metallography}
\bvolume{55}(\bissue{8}),
\bfpage{493}--\blpage{514}
(\byear{2018})
\doiurl{10.3139/147.110539}
\end{barticle}
\endbibitem

\bibitem[\protect\citeauthoryear{Adamovic et~al.}{2020}]{EMMC-MatDig}
\begin{botherref}
\oauthor{\bsnm{Adamovic}, \binits{N.}},
\oauthor{\bsnm{Friis}, \binits{J.}},
\oauthor{\bsnm{Goldbeck}, \binits{G.}},
\oauthor{\bsnm{Hashibon}, \binits{A.}},
\oauthor{\bsnm{Hermansson}, \binits{K.}},
\oauthor{\bsnm{Hristova‐Bogaerds}, \binits{D.}},
\oauthor{\bsnm{Koopmans}, \binits{R.}},
\oauthor{\bsnm{Wimmer}, \binits{E.}}:
The emmc roadmap for materials modelling and digitalisation of the materials
  sciences
(2020)
\doiurl{10.5281/zenodo.4272033}
\end{botherref}
\endbibitem

\bibitem[\protect\citeauthoryear{Wilkinson et~al.}{2016}]{wilkinson2016}
\begin{barticle}
\bauthor{\bsnm{Wilkinson}, \binits{M.D.}},
\bauthor{\bsnm{Dumontier}, \binits{M.}},
\bauthor{\bsnm{Aalbersberg}, \binits{I.J.}},
\bauthor{\bsnm{Appleton}, \binits{G.}},
\bauthor{\bsnm{Axton}, \binits{M.}},
\bauthor{\bsnm{Baak}, \binits{A.}},
\bauthor{\bsnm{Blomberg}, \binits{N.}},
\bauthor{\bsnm{Boiten}, \binits{J.-W.}},
\bauthor{\bsnm{Silva~Santos}, \binits{L.B.}},
\bauthor{\bsnm{Bourne}, \binits{P.E.}}, \betal:
\batitle{The fair guiding principles for scientific data management and
  stewardship}.
\bjtitle{Scientific data}
\bvolume{3}(\bissue{1}),
\bfpage{1}--\blpage{9}
(\byear{2016})
\end{barticle}
\endbibitem

\bibitem[\protect\citeauthoryear{McBride}{2004}]{mcbride2004resource}
\begin{botherref}
\oauthor{\bsnm{McBride}, \binits{B.}}:
The resource description framework (rdf) and its vocabulary description
  language rdfs.
Handbook on ontologies,
51--65
(2004)
\end{botherref}
\endbibitem

\bibitem[\protect\citeauthoryear{Bechhofer et~al.}{2004}]{bechhofer2004owl}
\begin{barticle}
\bauthor{\bsnm{Bechhofer}, \binits{S.}},
\bauthor{\bsnm{Van~Harmelen}, \binits{F.}},
\bauthor{\bsnm{Hendler}, \binits{J.}},
\bauthor{\bsnm{Horrocks}, \binits{I.}},
\bauthor{\bsnm{McGuinness}, \binits{D.L.}},
\bauthor{\bsnm{Patel-Schneider}, \binits{P.F.}},
\bauthor{\bsnm{Stein}, \binits{L.A.}}, \betal:
\batitle{Owl web ontology language reference}.
\bjtitle{W3C recommendation}
\bvolume{10}(\bissue{2}),
\bfpage{1}--\blpage{53}
(\byear{2004})
\end{barticle}
\endbibitem

\bibitem[\protect\citeauthoryear{P{\'e}rez et~al.}{2006}]{perez2006semantics}
\begin{bchapter}
\bauthor{\bsnm{P{\'e}rez}, \binits{J.}},
\bauthor{\bsnm{Arenas}, \binits{M.}},
\bauthor{\bsnm{Gutierrez}, \binits{C.}}:
\bctitle{Semantics and complexity of sparql}.
In: \bbtitle{The Semantic Web-ISWC 2006: 5th International Semantic Web
  Conference, ISWC 2006, Athens, GA, USA, November 5-9, 2006. Proceedings 5},
pp. \bfpage{30}--\blpage{43}
(\byear{2006}).
\bcomment{Springer}
\end{bchapter}
\endbibitem

\bibitem[\protect\citeauthoryear{Ihsan et~al.}{2023}]{IhsanDISO2023}
\begin{bchapter}
\bauthor{\bsnm{Ihsan}, \binits{A.Z.}},
\bauthor{\bsnm{Fathalla}, \binits{S.}},
\bauthor{\bsnm{Sandfeld}, \binits{S.}}:
\bctitle{Diso: A domain ontology for modeling dislocations in crystalline
  materials}.
In: \bbtitle{Proceedings of the 38th ACM/SIGAPP Symposium on Applied
  Computing}.
\bsertitle{SAC '23},
pp. \bfpage{1746}--\blpage{1753}.
\bpublisher{Association for Computing Machinery},
\blocation{New York, NY, USA}
(\byear{2023}).
\doiurl{10.1145/3555776.3578739}
\end{bchapter}
\endbibitem

\bibitem[\protect\citeauthoryear{Ihsan et~al.}{2021}]{ihsan2021_1}
\begin{bchapter}
\bauthor{\bsnm{Ihsan}, \binits{A.Z.}},
\bauthor{\bsnm{Dessi}, \binits{D.}},
\bauthor{\bsnm{Alam}, \binits{M.}},
\bauthor{\bsnm{Sack}, \binits{H.}},
\bauthor{\bsnm{Sandfeld}, \binits{S.}}:
\bctitle{Steps towards a dislocation ontology for crystalline materials}.
In: \bbtitle{2nd International Workshop on Semantic Digital Twins, SeDiT 2021},
vol. \bseriesno{2887}
(\byear{2021}).
\bcomment{CEUR-WS}
\end{bchapter}
\endbibitem

\bibitem[\protect\citeauthoryear{Berrueta et~al.}{2008}]{berrueta2008cooking}
\begin{bchapter}
\bauthor{\bsnm{Berrueta}, \binits{D.}},
\bauthor{\bsnm{Fern{\'a}ndez}, \binits{S.}},
\bauthor{\bsnm{Frade}, \binits{I.}}:
\bctitle{Cooking http content negotiation with vapour}.
In: \bbtitle{Proceedings of 4th Workshop on Scripting for the Semantic Web
  (SFSW2008)},
vol. \bseriesno{72}
(\byear{2008}).
\bcomment{Citeseer}
\end{bchapter}
\endbibitem

\bibitem[\protect\citeauthoryear{Loibl et~al.}{2020}]{loibl2020procedure}
\begin{barticle}
\bauthor{\bsnm{Loibl}, \binits{A.}},
\bauthor{\bsnm{Manoharan}, \binits{T.}},
\bauthor{\bsnm{Nagarajah}, \binits{A.}}:
\batitle{Procedure for the transfer of standards into machine-actionability}.
\bjtitle{Journal of Advanced Mechanical Design, Systems, and Manufacturing}
\bvolume{14}(\bissue{2}),
\bfpage{0022}--\blpage{0022}
(\byear{2020})
\end{barticle}
\endbibitem

\bibitem[\protect\citeauthoryear{Say et~al.}{2020}]{say2020semantic}
\begin{bchapter}
\bauthor{\bsnm{Say}, \binits{A.}},
\bauthor{\bsnm{Fathalla}, \binits{S.}},
\bauthor{\bsnm{Vahdati}, \binits{S.}},
\bauthor{\bsnm{Lehmann}, \binits{J.}},
\bauthor{\bsnm{Auer}, \binits{S.}}:
\bctitle{Semantic representation of physics research data}.
In: \bbtitle{Proceedings of the 12th International Joint Conference on
  Knowledge Discovery, Knowledge Engineering and Knowledge Management}
(\byear{2020})
\end{bchapter}
\endbibitem

\bibitem[\protect\citeauthoryear{Hu et~al.}{2011}]{hu2011agont}
\begin{bchapter}
\bauthor{\bsnm{Hu}, \binits{S.}},
\bauthor{\bsnm{Wang}, \binits{H.}},
\bauthor{\bsnm{She}, \binits{C.}},
\bauthor{\bsnm{Wang}, \binits{J.}}:
\bctitle{Agont: ontology for agriculture internet of things}.
In: \bbtitle{Computer and Computing Technologies in Agriculture IV: 4th IFIP TC
  12 Conference, CCTA 2010, Nanchang, China, October 22-25, 2010, Selected
  Papers, Part I 4},
pp. \bfpage{131}--\blpage{137}
(\byear{2011}).
\bcomment{Springer}
\end{bchapter}
\endbibitem

\bibitem[\protect\citeauthoryear{Say et~al.}{2020}]{say2020ontology}
\begin{bchapter}
\bauthor{\bsnm{Say}, \binits{Z.}},
\bauthor{\bsnm{Fathalla}, \binits{S.}},
\bauthor{\bsnm{Vahdati}, \binits{S.}},
\bauthor{\bsnm{Lehmann}, \binits{J.}},
\bauthor{\bsnm{Auer}, \binits{S.}}:
\bctitle{Ontology design for pharmaceutical research outcomes}.
In: \bbtitle{International Conference on Theory and Practice of Digital
  Libraries},
pp. \bfpage{119}--\blpage{132}
(\byear{2020}).
\bcomment{Springer}
\end{bchapter}
\endbibitem

\bibitem[\protect\citeauthoryear{Mrdjenovich
  et~al.}{2020}]{mrdjenovich2020propnet}
\begin{barticle}
\bauthor{\bsnm{Mrdjenovich}, \binits{D.}},
\bauthor{\bsnm{Horton}, \binits{M.K.}},
\bauthor{\bsnm{Montoya}, \binits{J.H.}},
\bauthor{\bsnm{Legaspi}, \binits{C.M.}},
\bauthor{\bsnm{Dwaraknath}, \binits{S.}},
\bauthor{\bsnm{Tshitoyan}, \binits{V.}},
\bauthor{\bsnm{Jain}, \binits{A.}},
\bauthor{\bsnm{Persson}, \binits{K.A.}}:
\batitle{Propnet: a knowledge graph for materials science}.
\bjtitle{Matter}
\bvolume{2}(\bissue{2}),
\bfpage{464}--\blpage{480}
(\byear{2020})
\end{barticle}
\endbibitem

\bibitem[\protect\citeauthoryear{Zhao et~al.}{2021}]{zhao2021knowledge}
\begin{bchapter}
\bauthor{\bsnm{Zhao}, \binits{X.}},
\bauthor{\bsnm{Greenberg}, \binits{J.}},
\bauthor{\bsnm{McClellan}, \binits{S.}},
\bauthor{\bsnm{Hu}, \binits{Y.-J.}},
\bauthor{\bsnm{Lopez}, \binits{S.}},
\bauthor{\bsnm{Saikin}, \binits{S.K.}},
\bauthor{\bsnm{Hu}, \binits{X.}},
\bauthor{\bsnm{An}, \binits{Y.}}:
\bctitle{Knowledge graph-empowered materials discovery}.
In: \bbtitle{2021 IEEE International Conference on Big Data (Big Data)},
pp. \bfpage{4628}--\blpage{4632}
(\byear{2021}).
\bcomment{IEEE}
\end{bchapter}
\endbibitem

\bibitem[\protect\citeauthoryear{Jain et~al.}{2013}]{jain2013}
\begin{barticle}
\bauthor{\bsnm{Jain}, \binits{A.}},
\bauthor{\bsnm{Ong}, \binits{S.P.}},
\bauthor{\bsnm{Hautier}, \binits{G.}},
\bauthor{\bsnm{Chen}, \binits{W.}},
\bauthor{\bsnm{Richards}, \binits{W.D.}},
\bauthor{\bsnm{Dacek}, \binits{S.}},
\bauthor{\bsnm{Cholia}, \binits{S.}},
\bauthor{\bsnm{Gunter}, \binits{D.}},
\bauthor{\bsnm{Skinner}, \binits{D.}},
\bauthor{\bsnm{Ceder}, \binits{G.}},
\bauthor{\bsnm{Persson}, \binits{K.a.}}:
\batitle{{The Materials Project: A materials genome approach to accelerating
  materials innovation}}.
\bjtitle{APL Materials}
\bvolume{1}(\bissue{1}),
\bfpage{011002}
(\byear{2013})
\doiurl{10.1063/1.4812323}
\end{barticle}
\endbibitem

\bibitem[\protect\citeauthoryear{Ashino}{2010}]{ashino2010}
\begin{barticle}
\bauthor{\bsnm{Ashino}, \binits{T.}}:
\batitle{Materials ontology: An infrastructure for exchanging materials
  information and knowledge}.
\bjtitle{Data Science Journal}
\bvolume{9},
\bfpage{54}--\blpage{61}
(\byear{2010})
\end{barticle}
\endbibitem

\bibitem[\protect\citeauthoryear{Li et~al.}{2020}]{li2020}
\begin{bchapter}
\bauthor{\bsnm{Li}, \binits{H.}},
\bauthor{\bsnm{Armiento}, \binits{R.}},
\bauthor{\bsnm{Lambrix}, \binits{P.}}:
\bctitle{An ontology for the materials design domain}.
In: \bbtitle{International Semantic Web Conference},
pp. \bfpage{212}--\blpage{227}
(\byear{2020}).
\bcomment{Springer}
\end{bchapter}
\endbibitem

\bibitem[\protect\citeauthoryear{McCusker et~al.}{2020}]{mccusker2020nanomine}
\begin{bchapter}
\bauthor{\bsnm{McCusker}, \binits{J.P.}},
\bauthor{\bsnm{Keshan}, \binits{N.}},
\bauthor{\bsnm{Rashid}, \binits{S.}},
\bauthor{\bsnm{Deagen}, \binits{M.}},
\bauthor{\bsnm{Brinson}, \binits{C.}},
\bauthor{\bsnm{McGuinness}, \binits{D.L.}}:
\bctitle{Nanomine: A knowledge graph for nanocomposite materials science}.
In: \bbtitle{International Semantic Web Conference},
pp. \bfpage{144}--\blpage{159}
(\byear{2020}).
\bcomment{Springer}
\end{bchapter}
\endbibitem

\bibitem[\protect\citeauthoryear{Ghoniem and Sun}{1999}]{ghoniem1999}
\begin{barticle}
\bauthor{\bsnm{Ghoniem}, \binits{N.M.}},
\bauthor{\bsnm{Sun}, \binits{L.}}:
\batitle{Fast-sum method for the elastic field of three-dimensional dislocation
  ensembles}.
\bjtitle{Physical Review B}
\bvolume{60}(\bissue{1}),
\bfpage{128}
(\byear{1999})
\end{barticle}
\endbibitem

\bibitem[\protect\citeauthoryear{Su{\'a}rez-Figueroa
  et~al.}{2015}]{suarez2015neon}
\begin{barticle}
\bauthor{\bsnm{Su{\'a}rez-Figueroa}, \binits{M.C.}},
\bauthor{\bsnm{G{\'o}mez-P{\'e}rez}, \binits{A.}},
\bauthor{\bsnm{Fernandez-Lopez}, \binits{M.}}:
\batitle{The neon methodology framework: A scenario-based methodology for
  ontology development}.
\bjtitle{Applied ontology}
\bvolume{10}(\bissue{2}),
\bfpage{107}--\blpage{145}
(\byear{2015})
\end{barticle}
\endbibitem

\bibitem[\protect\citeauthoryear{Simperl
  et~al.}{2011}]{Simperl2011-Ontology-metadata}
\begin{barticle}
\bauthor{\bsnm{Simperl}, \binits{E.}},
\bauthor{\bsnm{Sarasua}, \binits{C.}},
\bauthor{\bsnm{Ungrangsi}, \binits{R.}},
\bauthor{\bsnm{Bürger}, \binits{T.}}:
\batitle{Ontology metadata for ontology reuse}.
\bjtitle{International Journal of Metadata, Semantics and Ontologies}
\bvolume{6}(\bissue{2}),
\bfpage{126}--\blpage{145}
(\byear{2011})
\doiurl{10.1504/IJMSO.2011.046579}
\end{barticle}
\endbibitem

\bibitem[\protect\citeauthoryear{Hodgson et~al.}{2014}]{hodgson2014qudt}
\begin{botherref}
\oauthor{\bsnm{Hodgson}, \binits{R.}},
\oauthor{\bsnm{Keller}, \binits{P.J.}},
\oauthor{\bsnm{Hodges}, \binits{J.}},
\oauthor{\bsnm{Spivak}, \binits{J.}}:
Qudt-quantities, units, dimensions and data types ontologies.
USA Available http://qudt. org March
\textbf{156}
(2014)
\end{botherref}
\endbibitem

\bibitem[\protect\citeauthoryear{Fathalla et~al.}{2019}]{seo2019}
\begin{bchapter}
\bauthor{\bsnm{Fathalla}, \binits{S.}},
\bauthor{\bsnm{Vahdati}, \binits{S.}},
\bauthor{\bsnm{Lange}, \binits{C.}},
\bauthor{\bsnm{Auer}, \binits{S.}}:
\bctitle{Seo: A scientific events data model}.
In: \beditor{\bsnm{Ghidini}, \binits{C.}},
\beditor{\bsnm{Hartig}, \binits{O.}},
\beditor{\bsnm{Maleshkova}, \binits{M.}},
\beditor{\bsnm{Sv{\'a}tek}, \binits{V.}},
\beditor{\bsnm{Cruz}, \binits{I.}},
\beditor{\bsnm{Hogan}, \binits{A.}},
\beditor{\bsnm{Song}, \binits{J.}},
\beditor{\bsnm{Lefran{\c{c}}ois}, \binits{M.}},
\beditor{\bsnm{Gandon}, \binits{F.}} (eds.)
\bbtitle{The Semantic Web -- ISWC 2019},
pp. \bfpage{79}--\blpage{95}.
\bpublisher{Springer},
\blocation{Cham}
(\byear{2019})
\end{bchapter}
\endbibitem

\bibitem[\protect\citeauthoryear{Fathalla et~al.}{2023}]{fathalla2023upper}
\begin{bchapter}
\bauthor{\bsnm{Fathalla}, \binits{S.}},
\bauthor{\bsnm{Lange}, \binits{C.}},
\bauthor{\bsnm{Auer}, \binits{S.}}:
\bctitle{An upper ontology for modern science branches and related entities}.
In: \bbtitle{European Semantic Web Conference},
pp. \bfpage{436}--\blpage{453}
(\byear{2023}).
\bcomment{Springer}
\end{bchapter}
\endbibitem

\bibitem[\protect\citeauthoryear{Ehrig}{2006}]{ehrig2006ontology}
\begin{bbook}
\bauthor{\bsnm{Ehrig}, \binits{M.}}:
\bbtitle{Ontology Alignment: Bridging the Semantic Gap}
vol. \bseriesno{4}.
\bpublisher{Springer}, \blocation{???}
(\byear{2006})
\end{bbook}
\endbibitem

\bibitem[\protect\citeauthoryear{Noy et~al.}{2000}]{noy2000algorithm}
\begin{bchapter}
\bauthor{\bsnm{Noy}, \binits{N.F.}},
\bauthor{\bsnm{Musen}, \binits{M.A.}}, \betal:
\bctitle{Algorithm and tool for automated ontology merging and alignment}.
(\byear{2000})
\end{bchapter}
\endbibitem

\bibitem[\protect\citeauthoryear{Morgado et~al.}{2020}]{morgado2020}
\begin{botherref}
\oauthor{\bsnm{Morgado}, \binits{J.F.}},
\oauthor{\bsnm{Ghedini}, \binits{E.}},
\oauthor{\bsnm{Goldbeck}, \binits{G.}},
\oauthor{\bsnm{Hashibon}, \binits{A.}},
\oauthor{\bsnm{Schmitz}, \binits{G.J.}},
\oauthor{\bsnm{Friis}, \binits{J.}},
\oauthor{\bsnm{Baas}, \binits{A.}}:
Mechanical testing ontology for digital-twins: A roadmap based on emmo.
SeDiT 2020: Semantic Digital Twins 2020,
3
(2020)
\end{botherref}
\endbibitem

\bibitem[\protect\citeauthoryear{Lebo et~al.}{2013}]{lebo2013prov}
\begin{botherref}
\oauthor{\bsnm{Lebo}, \binits{T.}},
\oauthor{\bsnm{Sahoo}, \binits{S.}},
\oauthor{\bsnm{McGuinness}, \binits{D.}},
\oauthor{\bsnm{Belhajjame}, \binits{K.}},
\oauthor{\bsnm{Cheney}, \binits{J.}},
\oauthor{\bsnm{Corsar}, \binits{D.}},
\oauthor{\bsnm{Garijo}, \binits{D.}},
\oauthor{\bsnm{Soiland-Reyes}, \binits{S.}},
\oauthor{\bsnm{Zednik}, \binits{S.}},
\oauthor{\bsnm{Zhao}, \binits{J.}}:
Prov-o: The prov ontology. w3c recommendation.
World Wide Web Consortium
(2013)
\end{botherref}
\endbibitem

\bibitem[\protect\citeauthoryear{Po and Ghoniem}{2014}]{Po2014_1}
\begin{barticle}
\bauthor{\bsnm{Po}, \binits{G.}},
\bauthor{\bsnm{Ghoniem}, \binits{N.}}:
\batitle{A variational formulation of constrained dislocation dynamics coupled
  with heat and vacancy diffusion}.
\bjtitle{Journal of the Mechanics and Physics of Solids}
\bvolume{66},
\bfpage{103}--\blpage{116}
(\byear{2014})
\doiurl{10.1016/j.jmps.2014.01.012}
\end{barticle}
\endbibitem

\bibitem[\protect\citeauthoryear{Arsenlis et~al.}{2007}]{arsenlis2007enabling}
\begin{barticle}
\bauthor{\bsnm{Arsenlis}, \binits{A.}},
\bauthor{\bsnm{Cai}, \binits{W.}},
\bauthor{\bsnm{Tang}, \binits{M.}},
\bauthor{\bsnm{Rhee}, \binits{M.}},
\bauthor{\bsnm{Oppelstrup}, \binits{T.}},
\bauthor{\bsnm{Hommes}, \binits{G.}},
\bauthor{\bsnm{Pierce}, \binits{T.G.}},
\bauthor{\bsnm{Bulatov}, \binits{V.V.}}:
\batitle{Enabling strain hardening simulations with dislocation dynamics}.
\bjtitle{Modelling and Simulation in Materials Science and Engineering}
\bvolume{15}(\bissue{6}),
\bfpage{553}
(\byear{2007})
\end{barticle}
\endbibitem

\bibitem[\protect\citeauthoryear{Devincre et~al.}{2011}]{devincre2011}
\begin{barticle}
\bauthor{\bsnm{Devincre}, \binits{B.}},
\bauthor{\bsnm{Madec}, \binits{R.}},
\bauthor{\bsnm{Monnet}, \binits{G.}},
\bauthor{\bsnm{Queyreau}, \binits{S.}},
\bauthor{\bsnm{Gatti}, \binits{R.}},
\bauthor{\bsnm{Kubin}, \binits{L.}}:
\batitle{Modeling crystal plasticity with dislocation dynamics simulations: The
  ‘micromegas’ code}.
\bjtitle{Mechanics of Nano-objects}
\bvolume{1},
\bfpage{81}--\blpage{100}
(\byear{2011})
\end{barticle}
\endbibitem

\bibitem[\protect\citeauthoryear{Demirci et~al.}{2023}]{demirci}
\begin{botherref}
\oauthor{\bsnm{Demirci}, \binits{A.}},
\oauthor{\bsnm{Steinberger}, \binits{D.}},
\oauthor{\bsnm{Stricker}, \binits{M.}},
\oauthor{\bsnm{Merkert}, \binits{N.}},
\oauthor{\bsnm{Weygand}, \binits{D.}},
\oauthor{\bsnm{Sandfeld}, \binits{S.}}:
Statistical analysis of discrete dislocation dynamics simulations: initial
  structures, cross-slip and microstructure evolution.
Modelling and Simulation in Materials Science and Engineering
(2023)
\end{botherref}
\endbibitem

\bibitem[\protect\citeauthoryear{Motz et~al.}{2009}]{motz2009initial}
\begin{barticle}
\bauthor{\bsnm{Motz}, \binits{C.}},
\bauthor{\bsnm{Weygand}, \binits{D.}},
\bauthor{\bsnm{Senger}, \binits{J.}},
\bauthor{\bsnm{Gumbsch}, \binits{P.}}:
\batitle{Initial dislocation structures in 3-d discrete dislocation dynamics
  and their influence on microscale plasticity}.
\bjtitle{Acta Materialia}
\bvolume{57}(\bissue{6}),
\bfpage{1744}--\blpage{1754}
(\byear{2009})
\end{barticle}
\endbibitem

\bibitem[\protect\citeauthoryear{Fan et~al.}{2021}]{fan2021strain}
\begin{barticle}
\bauthor{\bsnm{Fan}, \binits{H.}},
\bauthor{\bsnm{Wang}, \binits{Q.}},
\bauthor{\bsnm{El-Awady}, \binits{J.A.}},
\bauthor{\bsnm{Raabe}, \binits{D.}},
\bauthor{\bsnm{Zaiser}, \binits{M.}}:
\batitle{Strain rate dependency of dislocation plasticity}.
\bjtitle{Nature communications}
\bvolume{12}(\bissue{1}),
\bfpage{1845}
(\byear{2021})
\end{barticle}
\endbibitem

\bibitem[\protect\citeauthoryear{Steinberger et~al.}{2019}]{steinberger2020}
\begin{barticle}
\bauthor{\bsnm{Steinberger}, \binits{D.}},
\bauthor{\bsnm{Song}, \binits{H.}},
\bauthor{\bsnm{Sandfeld}, \binits{S.}}:
\batitle{Machine learning-based classification of dislocation microstructures}.
\bjtitle{Frontiers in Materials}
\bvolume{6},
\bfpage{141}
(\byear{2019})
\doiurl{10.3389/fmats.2019.00141}
\end{barticle}
\endbibitem

\bibitem[\protect\citeauthoryear{Stricker
  et~al.}{2018}]{stricker2018dislocation}
\begin{barticle}
\bauthor{\bsnm{Stricker}, \binits{M.}},
\bauthor{\bsnm{Sudmanns}, \binits{M.}},
\bauthor{\bsnm{Schulz}, \binits{K.}},
\bauthor{\bsnm{Hochrainer}, \binits{T.}},
\bauthor{\bsnm{Weygand}, \binits{D.}}:
\batitle{Dislocation multiplication in stage ii deformation of fcc multi-slip
  single crystals}.
\bjtitle{Journal of the Mechanics and Physics of Solids}
\bvolume{119},
\bfpage{319}--\blpage{333}
(\byear{2018})
\end{barticle}
\endbibitem

\bibitem[\protect\citeauthoryear{Boettiger}{2018}]{rdflib}
\begin{botherref}
\oauthor{\bsnm{Boettiger}, \binits{C.}}:
Rdflib: A High Level Wrapper Around the Redland Package for Common Rdf
  Applications.
(2018).
\doiurl{10.5281/zenodo.1098478}
\end{botherref}
\endbibitem

\bibitem[\protect\citeauthoryear{Fathalla et~al.}{2019}]{fathalla2019eventskg}
\begin{bchapter}
\bauthor{\bsnm{Fathalla}, \binits{S.}},
\bauthor{\bsnm{Lange}, \binits{C.}},
\bauthor{\bsnm{Auer}, \binits{S.}}:
\bctitle{Eventskg: A 5-star dataset of top-ranked events in eight computer
  science communities}.
In: \bbtitle{European Semantic Web Conference},
pp. \bfpage{427}--\blpage{442}
(\byear{2019}).
\bcomment{Springer}
\end{bchapter}
\endbibitem

\bibitem[\protect\citeauthoryear{Tartir et~al.}{2005}]{tartir200}
\begin{botherref}
\oauthor{\bsnm{Tartir}, \binits{S.}},
\oauthor{\bsnm{Arpinar}, \binits{I.B.}},
\oauthor{\bsnm{Moore}, \binits{M.}},
\oauthor{\bsnm{Sheth}, \binits{A.P.}},
\oauthor{\bsnm{Aleman-Meza}, \binits{B.}}:
Ontoqa: Metric-based ontology quality analysis
(2005)
\end{botherref}
\endbibitem

\end{thebibliography}

\end{document}